\documentclass[11pt]{article}
\usepackage{times}
\usepackage{hyperref}
\usepackage[dvips, paper=letterpaper, top=1in, bottom=.75in, left=1in, right=1in, nohead, includefoot, footskip=.25in]{geometry}
\usepackage{booktabs} 
\usepackage[ruled]{algorithm2e} 

\SetAlFnt{\small}
\SetAlCapFnt{\small}
\SetAlCapNameFnt{\small}
\SetAlCapHSkip{0pt}
\IncMargin{-\parindent}

\usepackage[utf8]{inputenc}
\usepackage{amsmath, amssymb}
\usepackage{bm}
\usepackage{multirow,array}
\usepackage{amsthm}
\usepackage{comment}
\usepackage{color}
\usepackage{floatrow}
\usepackage{graphicx}
\usepackage{cleveref}
\usepackage{float}
\usepackage{natbib}
\usepackage{comment}
\usepackage{cleveref}
\usepackage{subfigure}
\usepackage{caption}

\newtheorem{theorem}{Theorem}[section]
\newtheorem{corollary}[theorem]{Corollary}
\newtheorem{claim}{Claim}[theorem]
\newtheorem{lemma}[theorem]{Lemma}
\newtheorem{definition}[theorem]{Definition}
\newtheorem{assumption}[theorem]{Assumption}
\newtheorem{remark}[theorem]{Remark}

\newcommand{\R}{\mathbb{R}}
\newcommand{\tcpa}{{ \mbox{tCPA} }}
\newcommand{\Rev}{{ \mbox{Rev} }}
\renewcommand{\hm}{\hat{\mu}}

\newcommand{\todo}[1]{{\color{red}{\textbf{[TODO: #1]}}}}

\newcommand{\andresnote}[1]{{\color{orange}{[Andres: #1]}}}

\newcommand{\arielnote}[1]{{\color{Ariel: cyan}{#1}}}
\newcommand{\Bb}{\bold{b}}

\newcommand{\compet}{\mathcal{C}^\mathcal{A}}
\newcommand{\ela}{\mathcal{E}}

\setcitestyle{authoryear}

\begin{document}

\title{Platform Competition in the Autobidding World}

\author{Gagan Aggarwal \footnote{gagana@google.com, Google Research} \\ \and Andres Perlroth \footnote{perlroth@google.com, Google Research} \\ \and Ariel Schvartzman \footnote{aschvartzman@google.com, Google Research} \\ \and Mingfei Zhao \footnote{mingfei@google.com, Google Research}}


\maketitle

\begin{abstract}

We study the problem of auction design for advertising platforms that face strategic advertisers who are bidding across platforms.
Each advertiser's goal is to maximize their total value or conversions while satisfying some constraint(s) across all the platforms they participates in. In this paper, we focus on advertisers with return-over-investment (henceforth, ROI) constraints, i.e. each advertiser is trying to maximize value while making sure that their ROI across all platforms is no less than some target value. An advertiser interacts with the platforms through autobidders -- for each platform, the advertiser strategically chooses a target ROI to report to the platform's autobidder, which in turn uses a uniform bid multiplier to bid on the advertiser’s behalf on the queries owned by the given platform. 

Our main result is that for a platform trying to maximize revenue, 
competition with other platforms is a key factor to consider when designing their auction. While first-price auctions are optimal (for both revenue and welfare) in the absence of competition, this no longer holds true in multi-platform settings. We show that there exists a large class of advertiser valuations over queries 
such that, from the platform's perspective, running a second price auction dominates running a first price auction. 

Furthermore, our analysis reveals the key factors influencing platform choice of auction format: (i) intensity of competition among advertisers, (ii) sensitivity of bid landscapes to an auction change (driven by advertiser sensitivity to price changes), 
 and (iii) relative inefficiency of second-price auctions compared to first-price auctions. 
\end{abstract}

\begin{titlepage}

\maketitle


\end{titlepage}

\section{Introduction}

Online advertisers often optimize their advertising campaigns
across different advertising platforms like Amazon, Bing, Google, Meta and TikTok to reach their audience effectively. Each platform, driven by its own goals, strategically designs its auction mechanism to sell ad space. While existing research provides valuable insights on auction design within individual platforms,
understanding how competition between platforms influences auction design remains largely unexplored.

This question becomes even more important with the increasing popularity of automated bidding where advertisers opt for target-ROI-based bidding algorithms aiming to maximize their value while keeping their return over investment 
(ROI) above a given target. If advertisers' true preferences are consistent with such target-based bidding, i.e. advertisers are indeed trying to maximize their value while ensuring that their ROI stays above some target value (such advertisers are sometimes referred to as {\em value-maximizing agents with ROI constraints}), then the auction design of one platform creates externalities for the other platforms through the constraints of the advertisers. 

In this paper, we study the question of auction design for competing platforms that face strategic advertisers bidding across the platforms. We assume that the advertisers interact with the platforms using autobidders, and that autobidders are using {\em uniform} bidding algorithms, i.e. the bid they set on each query (of the relevant platform) is proportional to the advertiser’s value for this query. 
~\footnote{Uniform bidding, also known in the literature as pacing bid, are well-studied algorithms in the literature as their simplicity and good performance make them appealing from a practical standpoint \citep{yuan_towards, balseiro_gur_2019, batenietal14}} When there are value-maximizing agents with ROI constraints who are bidding uniformly on a \emph{single} platform, the platform can maximize both welfare and revenue by using a first-price auction (FPA) {\citep{yuan_towards}}. In contrast, we show that in the presence of multiple platforms,  competition across platforms 
is a key factor to consider when choosing an auction format. 
Indeed, with sufficient competition, a platform often benefits from using a Second Price Auction (SPA) mechanism instead.\footnote{Observe that our result does not imply that a platform using FPA is acting as monopolist in the market -- there are other reasons why a platform would use FPA that our theory does not cover. For example, in Display Ads, first-price auctions are widely used because they are credible \citep{credibility_2020} and increase spending confidence of advertisers \citep{google21,marketing2021} as they are charged by their own bid.
}
The key intuition behind this is as follows: when an advertiser is optimizing across multiple platforms, it tries to equalize its marginal ROI across platforms. The auction format used on a platform has a direct impact on the bid landscape\footnote{The bid landscape for an advertiser is a mapping from bid to (cost, value), which indicates the spend and value an advertiser would get if she submits certain bid, when bids for others are fixed.} (and hence the marginal ROI) an advertiser faces on that platform. Hence, advertiser's bids depend on the auction choice of \emph{every} platform. As such, in choosing its auction format, each platform trades off the efficiency gain it might get from the use of one auction format with the higher bids it might get from the use of another auction format.

We uncover and quantify this trade-off in a stylized model with two platforms and two advertisers. Each revenue-maximizing platform owns a continuum of queries and chooses between SPA and FPA as the per-query auction mechanism to sell all their queries. We assume that platforms are symmetric in the sense that they own the same distribution of inventory, with possibly asymmetric market shares.
Each advertiser is a value-maximizing agent with a target 
ROI constraint that upper-bounds the average cost per unit value (conversion) they can pay across all the platforms they participate in. The platforms and advertisers play a sequential game as follows:
\begin{enumerate}
\item First, all platforms independently choose between SPA and FPA to sell their queries. 
\item Second, the advertisers strategically choose target ROIs to submit to a platform-specific 
autobidder who bids on behalf of the advertiser on the respective platform.\footnote{Alternatively, we can allow the advertisers to directly submit a uniform bid to each platform.}$^{,}$ \footnote{Note that there is a separate autobidder corresponding to each advertiser and platform pair.}
\item Each autobidder (corresponding to an advertiser and a platform) takes its target ROI input and computes bids using a uniform bidding strategy, i.e. the bid for each query is proportional to advertiser's value for the query.\footnote{Uniform bidding is the optimal bidding strategy among all bidding strategies when the underlying auction is truthful like SPA \citep{aggarwal_ashwin_mehta_www19}. For other auction formats, an advertiser could potential benefit by splitting their campaigns to induce a non-uniform bidding on the platform. Such strategic behavior is not modeled 
in this work.} It makes sure that the input target ROI is not exceeded.
\item Finally, the queries are allocated and payment accrues according to the chosen auction format.
\end{enumerate}
Advertisers' targets and per-query valuations are common knowledge to all agents in the game and we consider subgame perfect equilibrium as our solution concept.

We first solve the advertisers' subgame and characterize the optimal bidding when platforms have already declared the auction formats.
\Cref{th1:opt_bidding} shows that an advertiser's optimal target ROIs for different platforms are such that they obtain the same marginal cost on each of the platforms they participate in. We show that if a platform is using SPA as its auction mechanism, then the marginal cost on that platform is exactly equal to the bid the autobidder sets on the advertiser's behalf. On the other hand, if the platform is using FPA, then the marginal cost on that platform has an additional term that depends on the bid landscape the advertiser faces on the platform. We further show that since in equilibrium landscapes are endogenously generated by the other advertisers' bids, the difference between the marginal costs of FPA and SPA depends on the elasticity of advertisers’ valuations. The more elastic the valuations, the more the advertiser's report changes in reaction to the auction change of the platform. We refer the advertiser's report change to as \emph{bidding reaction} throughout the paper.

In Section~\ref{sec:necessary_condition} we show that the 
bidding reaction is an important factor in allowing (SPA, SPA) to be an equilibrium of the platforms' game. For this, we first consider a constrained model where advertisers use the same uniform bid 
across all platforms and, hence, may not be able to equalize marginals. In this case, Theorem~\ref{thm:uniform_bidding} shows that FPA is a weakly dominant strategy for each platform.\footnote{That is, the platform's revenue by using FPA is weakly greater than using SPA regardless of what auction format other platforms are using.}$^,$\footnote{In the paper, we show that our proof-technique generalises for general N platforms and M advertisers.} We next study the case where there is a single large advertiser facing a static bid landscape on each platform.\footnote{For example, this can happen when all the remaining advertisers are small value-maximizing advertisers that are interested in a single query.} In this case, the bidding reaction only comes from the large advertiser and hence the bidding reaction is weak. Theorem~\ref{thm:single-strategic} shows that FPA is a weakly dominant strategy for each platform in this case as well.

{In \Cref{sec:inefficiency_free} we study the inefficiency-free setting where, in equilibrium, queries are efficiently allocated regardless of the auction formats chosen by the platforms. 
}
We first consider the setting where both platforms have equal market share. Theorem~\ref{thm:symeq} shows that a platform's auction decision depends on (i) the elasticity of advertiser's valuations ($\ela$), which measures 
how strong the bidding reaction is and (ii) the relative difference of advertisers' valuations across the queries they compete in, captured by a competition metric ($\compet$) which measures how close the valuations are across advertisers. When $\ela\cdot \compet >1$, then SPA is a dominant strategy and hence (SPA, SPA) is the unique equilibrium of the game. Conversely, if $\ela\cdot \compet <1$ then FPA is a dominant strategy and hence (FPA, FPA) is the unique equilibrium of the game. Thus, given a fixed level of competition, if the bidding reaction is strong enough then SPA is an optimal auction to choose. Likewise, given a fixed level of bidding reaction, as the market becomes sufficiently competitive within advertisers then SPA is an optimal auction to choose. 
{Interestingly, we further show that the same necessary and sufficient condition can be applied to the more general setting with multiple ($n\geq 2$) platforms and asymmetric market share (Theorem~\ref{thm:market_share_sym}): For every platform, SPA is a dominant strategy if $\ela\cdot \compet >1$ and FPA is a dominant strategy if $\ela\cdot \compet<1$}.\footnote{Observe that even if one of the platforms owns the whole market, because SPA is also efficient, that platform is indifferent between FPA and SPA.} 
{In other words, this result shows that from a platform's perspective, the elasticity and level of competition (both independent of the platform's market share) outweigh the market share effects in this setting.}
A consequence of these two theorems is that the intensity of competition across platforms (which depends on the number of platforms and market share) 
can only impact the platform's auction decision when there is an efficiency loss when moving from FPA to SPA. 

In Section~\ref{sec:efficiencybidding}, we tackle the case where there is also an efficiency tradeoff between FPA and SPA. Given the technical challenges in characterizing the equilibrium for a general model, we analyze the model for two families of valuations that are tractable. These valuation families depend on a single parameter, $\alpha$, which affects the intensity of competition across advertisers, the elasticity of the advertiser, and the level of efficiency loss of SPA relative to FPA. By solving for the equilibrium, we find that the level of inefficiency of SPA vs. FPA also plays an important role in determining which auction format a platform should choose. In the first example, we observe that, even though elasticity and competition are both increasing in $\alpha$, FPA becomes the dominant strategy for large $\alpha$ as the inefficiency loss increases. In the second example, we observe that, for all $\alpha$, the inefficiency loss relative to the competition and elasticity is small and therefore, in equilibrium, it is a dominant strategy for platforms to use SPA. These examples showcase that for a general environment, there are three main factors that a platform should consider when choosing their auction format -- level of advertiser competition across queries, elasticity of advertisers, and the level of inefficiency of SPA compared to FPA.  


\subsection*{Practical Implications of Our Results}
As noted above, when there is only one platform facing value-maximizing advertisers with ROI constraints who are using uniform bidding (either directly or indirectly through autobidders), it is optimal for the platform to use FPA to maximize both welfare and revenue. This might motivate a platform to try using FPA even in the presence of other platforms. Our results show that advertiser response in the form of different target ROIs submitted to autobidders can often be large enough to affect which auction format is better for the platform. Thus, it is very important for a platform to properly account for advertiser response in evaluating different auction formats.

To make this point clear, suppose that a platform is currently using SPA and wants to experiment to see if switching to FPA would increase its revenue. We can then envision the following dynamics. Initially,  autobidders (that are typically algorithms that react rapidly to auction changes) would converge to a new bidding equilibrium -- for each advertiser, the autobidder will set a uniform bid multiplier equal to the current target (which is the same as the target submitted under SPA, $T^{SPA}$, since the advertiser has not yet responded -- see Theorem 6.1 \citet{yuan_towards}). After this first stage on the dynamics, the platform obtains a (weakly) higher revenue under FPA than SPA.\footnote{For a fixed set of targets within a platform, \citet{aggarwal_ashwin_mehta_www19} shows that the revenue of SPA can be half of the optimal welfare, while \citet{yuan_towards} shows that the revenue of FPA with uniform bidding is exactly the optimal welfare.} Thus, if the platform makes a decision at this stage of the dynamics, it would wrongly conclude that FPA is better. In the further stages of the dynamics, advertisers react when they observe that the value obtained from this platform is lower. This might lead them to decrease the target on this platform and increase the targets on other less expensive platforms. These adjustments would trigger iterative responses from auto-bidders across all platforms and further advertiser reactions. Our theory predicts that when this dynamic stabilizes, depending on the competition among advertisers, the elasticity of the advertiser's valuations and the SPA-to-FPA inefficiency trade-off, the final targets $T^{FPA}$ submitted by advertisers could be sufficiently low to make the FPA revenue lower than the SPA revenue. Thus, using measurements from an early stage of the dynamics could lead the platform to mistakenly adopt FPA.

\subsection{Related Work}
With the growth of autobidding in online ad auctions, there has been substantial research interest in problems related to autobidders.
We discuss some of the most relevant work and compare and contrast it with our results.
\paragraph{\bf Multi-Channel Mechanism design} \citet{aggarwal_perlroth_zhao_ec23} focuses on a single platform owning multiple channels, each selling their queries using SPA with a reserve price. The authors quantify the cost of having each channel optimizing its own reserve price compared to a global platform policy. In our paper, we study competition across platforms and consider the problem of platforms choosing between different auction mechanisms, not just optimizing their reserve prices. Motivated by the Display Ad market, \citet{renato_balu_yifeng_www2020} studies a model where multiple platforms compete for profit-maximizing bidders that are constrained to use the same bid on all platforms (in our language, they bid using a uniform bid).  Their main result shows that FPA is the optimal auction for the platforms, similar to \Cref{thm:single-strategic} where we show the same conclusion for the value-maximizing case.

\paragraph{\bf Bidding Algorithms for Autobidders, Uniform Bidding}
Uniform bidding (aka constant pacing) has been extensively studied in the context of budget constraints and, more generally, for Autobidding on a single platform. \cite{aggarwal_ashwin_mehta_www19} initiated the study of the Autobidding problem and, among other things, showed that uniform bidding is optimal if the platform uses a truthful auction format. \cite{10.1287/opre.2021.2167, doi:10.1287/mnsc.2022.4310} studies constant pacing for budget constraints on a platform that uses SPA or FPA. \cite{10.1145/2623330.2623366,10.1145/2783258.2788615} does an empirical study of the use of uniform bidding on real-world platforms when the bidder has budget constraints.
In this work we assume uniform bidding for autobidders and study the platform's auction choice in a multi-platform environment.

\paragraph{\bf Bidding across Multiple Platforms} Finally, \cite{susan2023multi} studies bidding strategies for utility-maximizing advertisers across multiple platforms under budget constraints. \cite{deng2023multi} focuses on bidding for value-maximizing advertisers under both budget and ROI constraints and shows that optimizing over per-platform ROIs can be arbitarily bad when the advertiser has both ROI and budget constraints. In contrast, with only ROI constraints, we show that marginal equalization using per-platform ROIs is the optimal bidding strategy.



\section{Model}
\label{sec:model}

\begin{figure}[h!]
    \centering
    \includegraphics[width=0.8\textwidth]{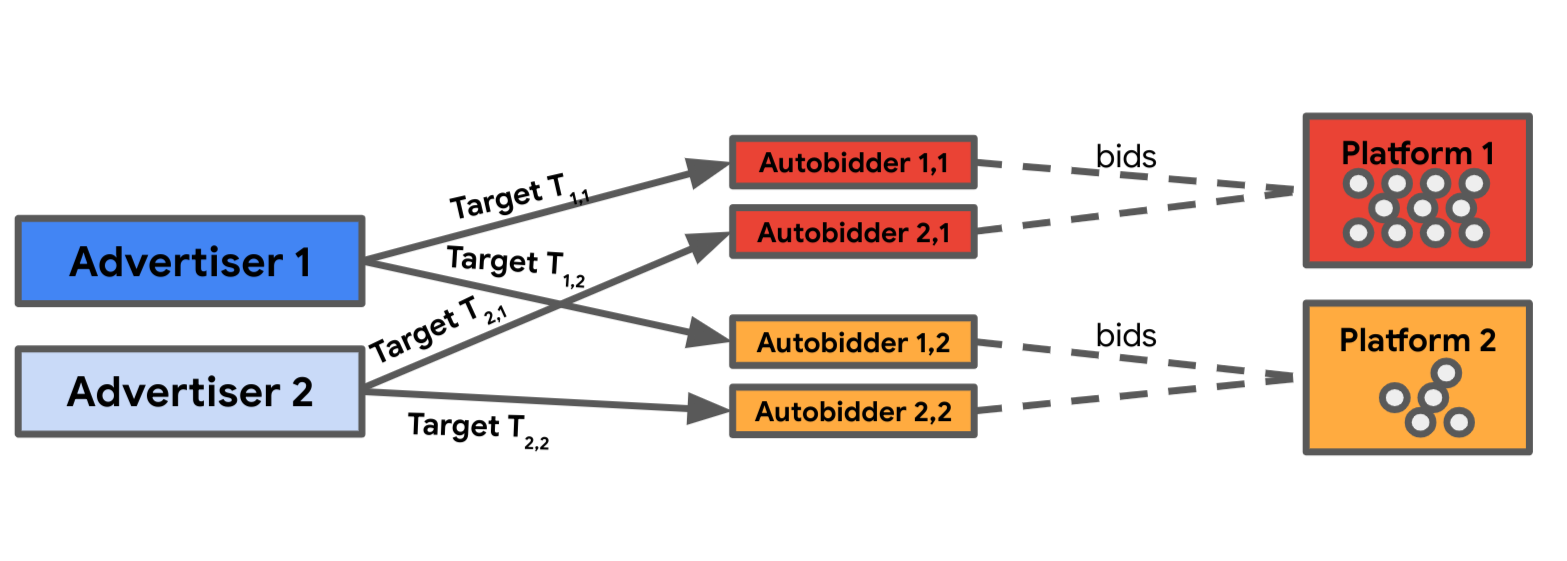}
    \caption{This figure captures the different agents (advertiser, platforms) involved in the game we are interested in studying. Advertiser $i$ submits target $T_{ij}$ to Autobidder $ij$, who bids on the their behalf on Platform $j$.}
    \label{fig:model}
\end{figure}

Our environment consists of two groups of strategic agents, platforms and advertisers, as well as autobidders who bid on an advertiser's behalf on different platforms (see Figure~\ref{fig:model}). 

\noindent{\bf Platforms:} 
There are $m$ platforms on the market and we denote $J=[m]$ the set of all platforms. Platform $j\in J$ owns a continuum of single-slot queries $q$ indexed on a set $Q_j$ which without of loss of generality we normalize to be $[0,1]$.\footnote{We use the standard Lebesgue measure and the Borelian $\sigma$-algebra for $[0,1]$} Each platform uses a per-query auction mechanism to sell the queries and would like to maximize its own revenue. Throughout this paper, we restrict the space of auction mechanisms to the two canonical auction rules: second-price auction (SPA) and first-price auction (FPA). 

\noindent{\bf Advertisers:}
There are $n$ advertisers interested in the queries and we denote $I=[n]$ the set of advertisers. Advertiser $i$'s valuation for query $q$ on platform $j$ is $v_{ij}(q)$, where $v_{ij}$ is a continuous differentiable function on $[0,1]$.\footnote{We consider the left derivative on $0$ and the right derivaitve on $1$} The advertiser's objective is to maximize the total value they get subject to having an average cost per value of at most a given target $T_i > 0$.\footnote{In this paper we focus on the setting where advertisers are ROI-constrained but not budget-constrained. It's an interesting open question to study the same problem with budget-constrained advertisers.} We notice that it is equivalent to having a target ROI of $1/T_i$. Denote the cost of winning query $q$ on platform $j$ by $c_{ij}(q)$. Then advertiser $i$'s objective consists on solving the following problem\footnote{All valuation functions are regular enough so that $v_{ij}(q)$ and $c_{ij}(q)$ are integrable functions.}
\begin{align*}
    \max_{(Q_{ij})} & \sum_{j\in J} \int_{Q_{ij}} v_{ij}(q)dq \\
    \mbox{s.t.} & \sum_{j\in J} \int_{Q_{ij}} c_{ij}(q)dq \leq T_i \cdot \sum_{j\in J} \int_{Q_{ij}} v_{ij}(q)dq.
\end{align*} 
Here we denote $Q_{ij}\subseteq Q_j$ the set of queries that advertiser $i$ wins on platform $j$. Observe that by re-scaling the value functions,
we assume without loss of generality that $T_i = 1$ for all $i\in I$. Advertiser $i$'s action consists of submitting a target $T_{ij}$ to the autobidder for each platform $j$, who bids on behalf of advertiser $i$ as described below.

\noindent{\bf Autobidders:}
For a given platform $j$, autobidder $ij$ receives as input target $T_{ij}$ from advertiser $i$ and bids $\mu_{ij}v_{ij}(q)$ on query $q\in Q_{j}$. The uniform bid multiplier $\mu_{ij}$ is computed by the autobidder for each advertiser independently with the goal to maximize advertiser $i$'s value on the platform:
\begin{align*}
    \max_{Q_{ij}} & \int_{Q_{ij}} v_{ij}(q)dq \\
    \mbox{s.t.} & \int_{Q_{ij}} c_{ij}(q)dq \leq T_{ij} \cdot \int_{Q_{ij}} v_{ij}(q)dq,
\end{align*} 
where $c_{ij}(q)$ is endogenously obtained as function of the auction format chosen by the platform and the bids received for query $q$.\footnote{For any query $q$ such that $\mu_{ij}v_{ij}(q)> \mu_{i'j}v_{i'j}(q), \forall i'\not=i$, advertiser $i$ wins query $q$. If the auction rule is SPA then $c_{ij}(q) = \max_{i'\not=i}\mu_{i'j}v_{i'j}(q)$. If the auction rule is FPA, the cost is instead $c_{ij}(q) = \mu_{ij}v_{ij}(q)$.}
 
\subsection*{Timing, Information and Equilibria}

The problem we are interested in studying is modeled by the following sequential game where platforms move first and advertisers respond. 
\begin{enumerate}
    \item Platform $j$ independently and simultaneously announces if it will use a SPA or FPA to sell their queries. 
    \item Advertiser $i$ independently and simultaneously submits targets $T_{ij}$ to the autobidder $ij$ managing platform $j$.
    \item For each Advertiser $i$, autobidder $ij$ computes an optimal bid multiplier $\mu_{ij}$, and bids $\mu_{ij}v_{ij}(q)$ for advertiser $i$ on query $q\in Q_j$ in platform $j$.
    \item Allocations and payoffs accrue according to the auction rules.
\end{enumerate}

We assume that the advertisers' valuations and targets are common knowledge to all players, so that the game is of complete information.~\footnote{This is a reasonable assumption in complex settings, see e.g.~\cite{yeganeh_mehta_perlroth_ec23}.} Our solution concept is subgame perfect Nash equilibrium (SPNE).

\begin{remark}
Given that the game is of complete information, in any SPNE, advertiser $i$ can foretell what bid multiplier autobidder $j$ will submit on her behalf. Thus, in the rest of the paper, we omit the autobidders' role and let the advertiser directly choose $\mu_{ij}$ without explicitly submitting $T_{ij}$ to autobidders.
\end{remark}

Another way to think about this is that the bidding interface provided by a platform allows submission of a $\mu_{ij}$ by advertiser $i$, with the understanding that $i$'s bid for query $q$ on platform $j$ is $\mu_{ij}v_{ij}(q)$.

We impose the following equilibrium refinements on the advertisers' bidding subgame to remove unnatural equilibrium and asymmetric equilibrium in symmetric environments. 
\begin{assumption}\label{as:refinments}
Given SPNE, let $(\mu_{ij})_{i\in I, j\in J}$ be the bidding strategies for a given history of the game. We impose the following refinements:
\begin{itemize}
    \item[(R1)] The equilibrium bid multiplier $\mu_{ij}$ is not weakly dominated by a bid multiplier $\mu'_{ij}$.
    \item[(R2)] If two advertisers have mirroring valuations on all platforms,\footnote{Two advertisers $i,i'$ have mirroring valuations on platform $j$ if $v_{ij}(q) = v_{i'j}(1-q)$ for all $q\in [0,1]$ (see \Cref{sec:inefficiency_free} for more details.)} advertisers use the same bid multiplier.  
    \item[(R3)] If each advertiser has the same valuation function on two different platforms, and both platforms use the same auction format (i.e. the two platforms are symmetric both in terms of valuations as well as auction format), then each advertiser uses the same bid multiplier on both platforms. 
\end{itemize}
\end{assumption}
Refinement (R1) is standard in the literature (e.g., \citep{aggarwal_ashwin_mehta_www19, aggarwal_perlroth_zhao_ec23}) and removes unnatural equilibria: for example, if there is a single platform running SPA, then the equilibrium where one bidder bids a large number and the second bidder bids $0$ does not pass this refinement as the second bidder's bid is dominated by bidding $1$. Refinements (R2)-(R3) restrict to symmetric bidding equilibrium whenever advertisers' valuations are symmetric in the mirroring sense (by reordering the queries) or platforms are fully symmetric in terms of their inventory and auction choice.

\subsection*{Benchmarks}

Before we finish this section, we define the total liquid welfare of the system as the allocation that maximizes total value across advertisers. This is also an upper-bound on the total revenue that can be generated across platforms.

\begin{definition}[Liquid Welfare]
The (optimal) liquid welfare of the system is defined as
$$ W^* = \max_{(Q_{ij})} \sum_{i\in I}\sum_{j\in J} \int_{Q_{ij}}v_{ij}(q)dq,$$
where $(Q_{ij})_{i\in I}$ is a partition of $Q_j$ for all $j\in J$.
\end{definition}

\begin{remark}\label{rem:LW}
Let $\Rev_j$ denote Platform $j$'s revenue in a given equilibrium. Then, $\sum_{j\in J} \Rev_J \leq W^* $.
\end{remark}

We introduce the competition benchmark $\compet \in [0, 1]$ {that} captures the level of advertiser competition within a platform. 

\begin{definition}[Competition]\label{def:competition}
Given advertisers valuations $(v_{ij})_{i\in I, j\in J}$, we define the level of advertisers' competition on platform $j$ by
\begin{equation*}
\compet_j = 1- \frac{\int_0^1 v_{(1)j}(q)-v_{(2)j}(q) dq}{\int_0^1 v_{(1)j}(q) dq},
\end{equation*}
where $v_{(1)j}(q), v_{(2)j}(q)$ are the highest and second-highest valuations for query $q$, respectively.
\label{def:competition}
\end{definition}

The competition metric $\compet_j$ measures, on average across the queries, the difference in valuation between the highest and second highest bidders. The metric is normalized by the liquid welfare so that it lies in $[0,1]$ and is scale-free. $\compet_j = 0$ implies that in each query there is only one relevant advertiser that can win the query, whereas if $\compet_j = 1$ for every query, there are multiple bidders with the same value for it. In particular, if there are just two bidders then their valuations for the queries are identical.

\section{The Advertiser Bidding Subgame: Marginal Cost Equalization }\label{sec:advertiser_subgame}

In this section, we study the bidding subgame between advertisers under a fixed choice of  auction formats for the platforms. {All proofs are deferred to \Cref{app:sec_optimal_bidding}.}

In Section~\ref{sec:adv-opt-bidding}, we first study a single advertiser's bidding problem, i.e. how to set bid multipliers on different platforms after observing the auction rules the platforms commit to implement and the others' bids. The main result in this section is that the advertisers will set their bid multipliers such that the marginal cost per unit value, henceforth marginal cost, is the same across all the platforms they participate in. We find the advertiser's best strategy by solving for the optimal bidding strategy when facing a general auction environment. We show that {when the bid landscapes, i.e. the mapping from multipliers to the total value and cost,} satisfy standard convexity conditions (see \Cref{theo:marg_sufficient_condition}), equalizing marginals across platforms is a necessary and sufficient condition for optimality.

In Section~\ref{sec:adv-bid-eq}, we study the bidding equilibrium of the subgame, that is when landscapes are endogenous, and {characterize}
the equilibrium conditions for the case of two symmetric platforms and two advertisers. We define an advertiser's {\em elasticity}, a key component that helps understand the bidding reaction advertiser's experience when a platform deviates from FPA to SPA.

\subsection{Advertiser's Optimal Bidding}
\label{sec:adv-opt-bidding}
We fix an advertiser $i$ and bids of the other advertisers. Denote by  $\Bb=(b_{i'j}(q))_{i'\not=i, j\in J, q\in Q_j}$ the bid functions for all other advertisers.
We denote by $V_{ij}(\mu_{ij}, \Bb)$ the aggregate value the advertiser obtains on platform $j$ when using a bid multiplier $\mu_{ij}$. Formally,
$V_{ij}(\mu_{ij}, \Bb) = \int_{Q_{ij}}v_{ij}(q)dq$, where $Q_{ij}=\{q\in Q_j: \mu_jv_{ij}(q)\geq b_{i'j}(q), \forall i'\not=i\}$ is the set of queries advertiser $i$ wins on platform $j$ by using bid multiplier $\mu_{ij}$. Throughout this section, we will simplify the notation and use $V_{j}(\mu_{j})$ instead when $i$ and $\Bb$ are fixed and clear from context.

Let $J_S,J_F\subseteq J$ be the set of platforms selecting SPA and FPA as their auction rule, respectively. Notice that from Myerson's characterization of truthful auctions, the total cost on platform $j\in J_S$ is given by $C_j(\mu_j)= \mu_j V_j(\mu_j) - \int_0^{\mu_j} V_j(z)dz$. On the other hand, for $j\in J_F$, the cost on the platform corresponds to the bid of $j$ on the queries they win, thus $C_j(\mu_j) = \mu_j V_j(\mu_j)$.
The advertiser's problem turns to
\begin{align}
    \max_{(\mu_j)_{j\in J}} &\sum_{j\in J} V_j(\mu_j) \label{bidder-problem}\\
    \mbox{s.t.}   &\sum_{j\in J} \mu_j V_j(\mu_j) - \sum_{j\in J_S}\int_0^{\mu_j} V_j(z)dz  \leq \sum_{j\in J} V_j(\mu_j) \label{eq:target-constraint}.
\end{align}

In what follows we impose the following conditions on the aggregate functions $(V_j)_{j\in J}$. 
\begin{assumption}\label{ass:vjs}
The functions $(V_j)_{j\in J}$ satisfy that
\begin{enumerate}
    \item $V_j(\mu_j)$ is increasing and twice-differentiable for $\mu \in [0, \hm_j]$ and that $V_j(\hm_j)$ equals a constant for $\mu > \hm_j $, for some $\hm_j \in (1,\infty]$.
    \item $V_j$ is concave.
\end{enumerate}
\end{assumption}
Assumption~\ref{ass:vjs} (1) implies that a bid level $\hm_j$ exists (potentially infinity) so that the advertiser gets all the inventory on the platform. Yet the advertiser needs to bid strictly more than their value (i.e. $\mu>1$) to buy all the inventory of platform $j$. 
Assumption~\ref{ass:vjs} (2) means that the marginal gain on value is decreasing as a function of the bid. We show in \Cref{lem0:marginals} that the marginal cost function is decreasing if $V_j$ is concave.

The following theorem provides the regularity conditions for existence of a solution to the advertiser problem.

\begin{theorem}\label{th1:opt_bidding}
Suppose that Assumption~\ref{ass:vjs} holds and if for some platform $j\in J_S$ we have that $\hm_j = \infty$ then
\begin{equation}\label{eq:condtion_hm_infty}
    \sum_{j\in J} \hm_j V_j(\hm_j) - \sum_{j\in J_S}\int_0^{\hm_j} V_j(z)dz > \sum_{j\in J} V_j(\hm_j).
\end{equation} 
Then, a solution $(\mu^*_j)_{j \in J}$ exists to Problem~\eqref{bidder-problem}. 
\end{theorem}
Observe that Condition~\eqref{eq:condtion_hm_infty} removes the case where the solution requires the advertiser to bid infinity on a SPA platform. The condition implies that buying all the inventory across the platforms violates the target CPA constraint (Equation~\eqref{eq:target-constraint}). 

\subsection*{Marginal Cost Equalization}

We now formalize the concept of {\em marginal cost} for this environment and show that it is a key element to characterize the advertiser's bidding problem. The marginal cost intuitively corresponds to the additional cost an advertiser would have to pay to marginally increase their value on a platform. Formally,
$$ MC_j(\mu_j) = \lim_{\Delta \to 0} \frac{C_j(\mu_j + \Delta) - C_j(\mu_j)}{V_j(\mu_j + \Delta) - V_j(\mu_j)}.$$

The following lemma shows that the marginal cost is directly related to the auction rule chosen by a platform.
\begin{lemma}\label{lem0:marginals}
Suppose that Assumption~\ref{ass:vjs} holds. Then for $\mu_j\in [0, \hm_j]$,\footnote{We consider the interior derivative for the extreme points $0,\hm_j$.}
$$MC_j(\mu_j)= \begin{cases}
\mu_j &\mbox{ if } j \in J_S\\
\mu_j + \frac{V_j(\mu_j)}{V_j'(\mu_j)} &\mbox{ if } j \in J_F
\end{cases}$$
Moreover, when $V_j$ is concave, we have that $MC_j(\mu_j)$ is an increasing function on $[0,\hm_j]$.
\end{lemma}

Observe that the marginal cost on an SPA platform ($j \in J_S$) is independent of 
the bids of other advertisers and only depends on the bid multiplier chosen by the advertiser. On the other hand, on an FPA platform, the marginal cost includes a second positive term which depends on other advertisers' bids. {This makes the marginal cost at a given bid larger on FPA than on SPA. }

To see how the marginal cost function is computed in \Cref{lem0:marginals}, suppose the advertiser increases their bid multiplier from $\mu_j$ to $\mu_j+\Delta$ so they get an increase in value of $ V_j(\mu_j+\Delta)-V_j(\mu_j)$. Because the auction is SPA, the change in cost by raising the bid multiplier by $\Delta$ only affects the new queries the bidder wins at $\mu_j+\Delta$. Thus, the change in cost is $\mu_j ( V_j(\mu_j+\Delta)-V_j(\mu_j)) + o(\Delta)$. Taking $\Delta\to 0$, this leads to a marginal cost of $\mu_j$. When the auction is FPA, the cost increase from increasing the bid multiplier to $\mu_j+\Delta$ not only comes from the new queries the advertiser obtains, but also from the increased payment for the old queries that it is already winning. This adds an extra cost of $\Delta V_j(\mu_j)$, leading to an excess term in the marginal cost of $\Delta V_j(\mu_j) /(V_j(\mu_j+\Delta)-V_j(\mu_j))= V_j(\mu_j)/V'_j(\mu_j)$ as $\Delta\to 0$.

The next theorem shows if the marginal costs are increasing, then an advertiser that equalizes marginal costs while meeting its target constraint bids optimally.

\begin{theorem}\label{theo:marg_sufficient_condition}
Suppose that Assumption~\ref{ass:vjs} holds. Then, if the system of equations
\begin{equation}\label{eq:marg_equalization}
\begin{cases}
MC_j(\mu_j)  = MC_{j'}(\mu_{j'}) \quad \mbox{ for } j,j' \in J\\
\sum_{j\in J} \mu_j V_j(\mu_j) - \sum_{j\in J_S}\int_0^{\mu_j} V_j(z)dz  = \sum_{j\in J} V_j(\mu_j)
\end{cases}
\end{equation}
has a nonzero solution $(\mu^*_j)_{j\in J}$, then this is the unique solution of Equation~\eqref{eq:marg_equalization} and the unique solution to Problem~\eqref{bidder-problem}.
\end{theorem}

We conclude this section by providing a sufficient condition on $V_j$ to have a non-trivial solution to~\eqref{eq:marg_equalization}. Therefore, we show that equalizing marginal costs is a necessary and sufficient condition to solve the advertiser's bidding problem. In \Cref{sec:adv-bid-eq}, we translate the condition into conditions on the advertisers' valuations.

\begin{theorem}\label{theo:sufficient_condition_general_statement}
Suppose that Assumption~\ref{ass:vjs} holds, and there exists a solution $(\mu^*_j)_{j\in J} $ to Problem~\eqref{bidder-problem} such that $\mu^*_j < \hm_j$ for all $j\in J$. Then,  $(\mu^*_j)_{j\in J} $ is the unique solution to Problem~\eqref{bidder-problem} and also the unique solution to~\eqref{eq:marg_equalization}.
\end{theorem}

\begin{corollary}\label{coro:vj_conditions}
Suppose Assumption~\ref{ass:vjs} holds with $\hm_j = \infty$ for all $j\in J$ and there is at least one platform using $FPA$. Then, the unique non-zero solution to~\eqref{eq:marg_equalization} is the unique solution to Problem~\eqref{bidder-problem}.
\end{corollary}

\subsection{Advertiser Bidding Equilibrium}
\label{sec:adv-bid-eq}
We now apply the previous results that characterize the best response of the advertiser to solve the advertisers' equilibrium in every platform subgame. We focus on instances where there are two symmetric platforms $J=\{a,b\}$ and two advertisers $I=\{1,2\}$. Given that platforms are symmetric we drop the index $j$ on advertisers' valuations.

Notice that by re-indexing the queries, without loss of generality we can assume that $h(q) =\frac{v_{1}(q)}{v_{2}(q)}$ is non-decreasing on $q$ so that queries are ordered non-decreasingly in terms of their value for advertiser 1 relative to advertiser 2. In other words, if advertiser 1 is getting query $q$, then advertiser 1 is also getting query $q'>q$. Moreover, if $h$ is an increasing function then for any non-zero bid multipliers $\mu_1,\mu_2$ used on a platform we have that $q(\mu_1,\mu_2) = \min_{q\in[0,1]}\{\mu_1 v_1(q) = \mu_2 v_2(q)\}$ is the query-threshold for such bid multipliers.\footnote{In case that the minimum does not exist, we take $q(\mu_1,\mu_2)=0$.} Thus, advertiser 1 wins all queries in $(q(\mu_1,\mu_2),1]$ and advertiser 2 gets queries $[0,q(\mu_1,\mu_2))$.\footnote{Because queries are a continuous set, it is irrelevant who wins query $q(\mu_1,\mu_2)$.} In addition, we define $q_{\text{eff}} =q(1,1) $ to be the threshold query in the efficient allocation.

We are now in position to solve the three possible subgames that can arise in our game: (FPA, FPA), (SPA, SPA) or (SPA, FPA).~\footnote{Due to the symmetry of the platforms, the solutions under (SPA, FPA) and (FPA, SPA) are identical.}\medskip

\noindent{\bf Case I: (FPA, FPA) subgame.}  Since platforms are symmetric and using the same auction format, our equilibrium refinement (see \Cref{as:refinments}) imposes that advertisers submit the same bid multipliers in both platforms. Hence, the subgame is equivalent to the case where there is a single platform using FPA. As discussed in \citet{yuan_towards}, a bid multiplier $\mu_i = 1$ is a dominant strategy for each of the advertisers, and therefore the equilibrium allocation is $q_{\text{eff}}$. We summarize these findings in the following theorem.

\begin{theorem}[(FPA, FPA) subgame \citep{yuan_towards}]\label{theo:fpa_subgame}
For the (FPA, FPA) subgame, the unique equilibrium with undominated strategies is that advertisers bid $\mu^*_i = 1$ for both platforms. The equilibrium query threshold $q_{\text{eff}}$ and the revenue each platform obtains $W^*/2$. 
\end{theorem} \medskip

\noindent{\bf Case II: (SPA, SPA) subgame.} As in the previous case, we have that the problem is equivalent to that of a single platform running SPA. The existence and uniqueness of equilibrium has been analyzed in \citet{yeganeh_mehta_perlroth_ec23}.

\begin{theorem}[(SPA,SPA) subgame \citep{yeganeh_mehta_perlroth_ec23}]\label{theo:spa_subgame}
Suppose that $h(q)$ is an increasing function. There exists an equilibrium and the equilibrium query-threshold $q_S = q(\mu^*_1,\mu^*_2) $ solves the equation    
\begin{align}
     \frac {\int_{q_S}^1 v_1(z)dz}{\int_{0}^{q_S}v_2(z)dz} = h(q_S) \cdot   \frac {\int_{q_S}^1 v_2(z)dz}{\int_{0}^{q_S}v_1(z)dz}.
    \end{align}
Moreover, if $h$ is convex and $v_2$ is decreasing on $[0,1]$, the equilibrium is unique.
\end{theorem}\medskip

\noindent{\bf Case III: (FPA, SPA) subgame.} In this case platforms are no longer symmetric in regards to the auction format and, thus, our subgame analysis can no longer be reduced to a single platform problem. For each advertiser $i$, denote $\mu_i^S$ and $\mu_i^F$ the bid multipliers of advertiser $i$ for the platform that uses SPA and FPA respectively.

The following lemma provides the conditions on the bid landscapes so that marginal cost equalization of both advertisers becomes a sufficient condition to pin-down equilibrium.

\begin{lemma}\label{lem:subgame_fpa_spa}
Suppose that $v_1$ is an increasing function, $v_2$ a decreasing function and $h$ is a convex function. Then for any positive bid multiplier $\mu_1$ ($\mu_2$) the landscape advertiser 2 (advertiser 1) faces 
$$V_{2j}(\mu) = \int_0^{g(\frac \mu {\mu_1})}v_2(q) dq \mbox{ (with } g = h^{-1}) \quad \left (V_{1j}(\mu) = \int_{f(\frac \mu {\mu_2})}^1 v_1(q) dq \mbox{ (with } f = (1/h)^{-1})\right)$$ satisfies \Cref{ass:vjs}.
\end{lemma}

We now use \Cref{lem0:marginals} to derive the condition for marginal cost equalization for an advertiser. For this goal, we first define an advertiser's {\em elasticity} -- a key metric that dictates how the marginal cost on the FPA platform compares to that on an SPA platform for a given bid. It is a combination of the additional cost paid by winning an additional query and the additional cost paid on existing won queries when the bid is increased to win an additional query. By applying the marginal equalization lemma,
we will show that elasticity dictates how different the bid multipliers of an advertiser are on FPA and SPA platforms. 

\begin{definition}[Elasticity] Let $\eta_1(q) = |{\frac{v'_1(q)}{v_1(q)}}|$. We define the elasticities for $v_1,v_2$ at query $q$   
\begin{align}
\ela_i(q)=1 + \left(\eta_1(q) + \eta_2(q)\right) \int_{q}^1 \frac {v_1(z)} {v_1(q)} dz  \\
\ela_2(q) = 1 + \left(\eta_1(q) + \eta_2(q)\right) \int_{0}^{q} \frac{v_2(z)}{v_2(q)} dz.
\end{align}
\end{definition}

\begin{lemma}\label{lem:marginal_cost_lem:fpa_spa}
Consider bid multipliers $(\mu_{i'}^F,\mu_{i'}^S)$ of advertiser $i'$. Then, the marginal cost of advertiser $i$ on the SPA platform is $MC_{i}^S(\mu_{i}^S) = \mu_{i}^S$ and on the FPA platform is $MC_{i}^F(\mu_{i}^F) = \mu_{i}^F\cdot \ela_{i}(q(\mu_{i}^F, \mu_{i'}^F))$.
\end{lemma}



We are now in position to present the main result of the section.
\begin{theorem}[(FPA, SPA) subgame: Marginal Cost Equalization]\label{theo:spa_fpa__subgame}
Suppose that $v_1$ is an increasing function, $v_2$ a decreasing function and $h$ is a convex function. Suppose $(\mu^{*F}_1, \mu^{*S}_1, \mu^{*F}_2,\mu^{*S}_2)$ is non-zero bid multiplier solution to the system of equations
    \begin{align}
    &\frac{\mu_1^S}{\mu_1^F} = \ela_1(q_F), \quad \qquad \;
    \frac{\mu_2^S}{\mu_2^F} = \ela_2(q_F), \label{eq:intrabidder1}  \\
      &      \mu_1^F \int_{q_F}^1 v_1(q)dq + \mu_2^S \int_{q_S}^1 v_1(q)dq = \int_{q_F}^1 v_1(q)dq + \int_{q_S}^1 v_1(q)dq,  \label{eq:tcpa-final-spa-fpa-2} \\
       & \mu_2^F \int_0^{q_F} v_2(q)dq + \mu_1^S \int_0^{q_S} v_1(q)dq = \int_0^{q_F} v_2(q)dq + \int_0^{q_S} v_2(q)dq  \label{eq:tcpa-final-spa-fpa}.
    \end{align}
    where $q_S,q_F$ are the equilibrium query thresholds, i.e., ${\mu_1^S}/{\mu_2^S} = h(q_S)$ and
       ${\mu_1^F}/{\mu_2^F} = h(q_F)$. Then $(\mu^{*F}_1, \mu^{*S}_1, \mu^{*F}_2,\mu^{*S}_2)$ conform an advertiser equilibrium of the subgame.
       
       Furthermore, if the above system of equations has a unique nonzero solution $(\mu^{*F}_1, \mu^{*S}_1, \mu^{*F}_2,\mu^{*S}_2)$ and $v_1(0) = v_2(1) = 0 $, the equilibrium is unique.
\end{theorem}

Observe that the system of equations \eqref{eq:intrabidder1}-\eqref{eq:tcpa-final-spa-fpa} correspond to the marginal cost equalization problem of \Cref{theo:marg_sufficient_condition} (Equation~\eqref{eq:marg_equalization}) which implies that a nontrivial solution is an equilibrium. To guarantee uniqueness of equilibrium we need to rule out the corner-case equilibrium where one of the advertiser is obtaining all the inventory of any platform and might not be equalizing their marginal costs. However, since $v_1(0) = v_2(1) = 0$ for any positive bid multiplier, advertiser 1 (or advertiser 2) cannot get a subset of queries close enough to $0$ (or $1$). Thus, with this assumption, there is no corner-case equilibrium.

\section{Importance of Bidding Reaction}
\label{sec:necessary_condition}

In this section, we present two examples that showcase the importance of the advertisers' bidding reaction in allowing for (SPA, SPA) to be a feasible equilibrium of the platform-advertiser game. When a platform changes its auction, the advertisers will solve the problem in Equation~(\ref{bidder-problem},\ref{eq:target-constraint}) and may change their bid multipliers in reaction to the auction change. We refer to this effect as the advertiser's \emph{bidding reaction}. 

In the first example, advertisers are {\em forced} to use the same bid multiplier across platforms, and thus cannot equalize marginal costs;\footnote{If all platforms are using SPA, notice from Theorem~\ref{th1:opt_bidding} that using the same bid multiplier is optimal, and hence, it equalizes the marginal cost.} In the second example, there is a single {\em large} advertiser bidding across platforms and competitors are {\em small} value-maximizing advertisers bidding in a single query, and thus, they don't adjust their bids when the auction changes. These two examples disable advertisers' reaction to the auction change to some degree. The main conclusion of this section is that in these two examples the platforms will always end up at the (FPA, FPA) equilibrium. We attribute this result in large part to the lack of advertisers' bidding reaction.


\subsection{Case I: Suboptimal uniform bidding across platforms}



We first study the case where advertisers are forced to use the same bid multiplier across all platforms, i.e. advertiser $i$'s bid on a query $q$ in platform $j$ is given by $\mu_i v_{ij}(q)$ for all $j \in J$.

The main result of this section is that if advertisers are using {\em weakly undominated}~\footnote{Weakly undominated strategies are those that are not weakly dominated by any other strategy.} bidding strategies then in the unique Nash equilibrium of the game, each platform uses FPA to sell their queries and in return, each advertiser uses a bid multiplier of $1$ across all platforms. Removing weakly dominated strategies, those where there are weakly dominant alternatives, precludes spurious equilibria. Furthermore, the revenue across all platforms matches the liquid welfare $W^*$. In particular, this implies that if a single entity owns all platforms, then the platform is implementing the revenue-maximizing policy. The proof of \Cref{thm:uniform_bidding} is deferred to \Cref{sec:unif_mult_across_plat}.


\begin{theorem}\label{thm:uniform_bidding}
The unique equilibrium of the platform-advertiser game under uniform bidding is for the platforms to announce FPA to sell their queries and for advertisers to submit multipliers $\mu_i = 1$. Furthermore, the total revenue across all platforms is equal to the liquid welfare $W^*$.
\end{theorem}


\subsection{Case II: Single Strategic advertiser vs. static advertisers}


In this section, we study the setting where a single strategic advertiser bids across all platforms. For each platform $j$ and each query $q$, the strategic advertiser competes with a static value-maximizing advertiser that only bids on this query. Since the static advertiser only participates in the auction for a single query, it is optimal for them to bid truthfully.\footnote{If the auction is FPA, their value is the largest bid to without violating the ROI constraint; Suppose the auction is SPA. When the second-highest bid is less than their value, bidding truthfully can win the auction without violating the ROI constraint. On the other hand, when the second-highest bid is larger than their value, they can not win without violating the constraint.} In other words, no static advertiser will respond to the auction change.

We show that for this case, the unique Nash equilibrium of the game is for all platforms to announce FPA to sell their queries, and for the strategic advertiser to submit a multiplier of $\mu_i = 1$.

\begin{theorem}\label{thm:single-strategic}
In the case of a single strategic advertiser against static advertisers, the unique Nash equilibrium is for the platforms to use FPA and for the strategic advertiser to use a multiplier of 1 on each platform.
\end{theorem}

The high-level idea of the proof is as follows. Since static advertisers will always bid their value regardless of the auction choice, the single strategic advertiser is competing with some static value function $v_j(\cdot)$ for each platform $j$. We notice that if a platform $j$ uses FPA, the revenue must be at least $\int_0^1 v_j(q)dq$. If the strategic advertiser wins some query $q$, their bid must be at least $v_j(q)$ which is also the payment; if the static advertiser wins, the payment is $v_j(q)$. Similarly, the revenue when platform $j$ uses SPA is at most $\int_0^1 v_j(q)dq$. Thus FPA is a weakly dominant strategy for all platforms. Finally, if all platforms use FPA, then it's clearly optimal for the advertiser to send their target on all platforms, i.e. use a bid multiplier of 1. See \Cref{sec:single-bidder} for a formal proof of \Cref{thm:single-strategic}.

\section{The Inefficiency-free Case}\label{sec:inefficiency_free}

In this section, we study the \textit{inefficiency-free} setting where queries are efficiently allocated at the bidding equilibrium of any auction outcome. Recall that $q_{\text{eff}} \in (0,1)$ is the threshold query such that $v_1(q_{\text{eff}})=v_2(q_{\text{eff}})$, which allocates efficiently. 
In this section we study settings where, for any profile of the platforms' auctions, the allocation is efficient at the (advertisers') bidding equilibrium, i.e. $q_{\text{eff}}$ is the allocation threshold in all platforms.

 
A simple example of the setting mentioned above is when the valuation functions of both advertisers are \textit{mirrored} versions of each other, i.e., $v_{1}(q)=v_{2}(1-q)$ for all queries $q\in [0,1]$, for some monotone increasing $v_{1}(\cdot)$. In this case, the advertisers are symmetric, and as noted in \Cref{sec:model}, we will focus on symmetric equilibria where where both advertisers use the same bid multiplier for each platform (the bid multipliers are the same across advertisers, but could vary by platform if they use different auction formats). Hence for all platforms, advertiser 1 wins $q\in [0,1/2]$ while advertiser 2 wins $q\in (1/2,1]$, making the allocation efficient regardless of the chosen auction format. 



We first solve the case where there are two symmetric platforms with equal market share competing for the advertisers. 
{We provide a full characterization of the platform equilibrium. In particular, we show that for any valuation function, either SPA or FPA is a dominant strategy for each platform. We give necessary and sufficient conditions under which each of FPA and SPA are dominant.} Then we extend our result to the case of multiple platforms with different market share. We show that the same conditions apply regardless of the number of platforms nor the market share. All proofs in this section are deferred to \Cref{sec:proof_inefficiency_free}. 


\subsection{Two Symmetric Platforms with Equal Market Share}\label{subsec:sym}






We first consider the case where platforms are symmetric (with equal market share). We show in \Cref{thm:symeq} a full characterization of the platform equilibrium of the game. 

We show that in this setting the elasticity of both advertisers are equal at $q_{\text{eff}}$, i.e. $\ela_1(q_{\text{eff}})=\ela_2(q_{\text{eff}})$ and simply denote it as $\ela$. Recall that $\compet=1-\frac{\int_0^1 |v_1(q)-v_2(q)|dq}{\int_0^1 \max\{v_1(q), v_2(q)\}dq}$ is the competition level of the game. Here we drop the subscript $j$ since platforms are symmetric. We prove the following theorem that characterizes the platform equilibrium of the game.

\begin{theorem}\label{thm:symeq}
For any inefficiency-free instance of 2 advertisers and 2 symmetric platforms, where $\ela = \ela_1(q_{\text{eff}}) = \ela_2(q_{\text{eff}})$ the following holds. If $\ela\cdot \compet > 1$, SPA is a dominant strategy and thus (SPA, SPA) is the only NE of the game. If $\ela\cdot \compet < 1$, then FPA is a dominant strategy of the game and (FPA, FPA) is the only NE. If $\ela\cdot \compet = 1$, then the game is degenerate: all four profiles of the game have the same outcome. 
\end{theorem}

\Cref{thm:symeq} implies that in this 
setting, the equilibrium of the game is only determined by the following two factors: the competition level of the game $\compet$ and the elasticity of the advertisers $\ela$. An instance with a larger competition level $\compet$ indicates enhanced competition between the advertisers, which makes SPA more favorable. As discussed in \Cref{sec:advertiser_subgame}, the elasticity indicates how large the bid change is when a platform changes it auction. \Cref{thm:symeq} provides a direct contrast with Theorems~\ref{thm:uniform_bidding},~\ref{thm:single-strategic}: given the competition level of the game, if the elasticity of the advertisers is sufficiently high, then SPA will be the dominant strategy of the game.



{We then examine the classic monomial value function $q^\alpha$ ($\alpha\geq 1$) in the mirrored setting, i.e. $v_1(q)=q^\alpha, v_2(q)=(1-q)^\alpha$. First we notice that $\frac{v_1(q)}{v_2(q)}=(\frac{q}{1-q})^\alpha$ is convex when $\alpha\geq 1$.} Using Theorem~\ref{thm:symeq}, we show that SPA is a dominant strategy for any monomial function. We examine other classic value functions and show that SPA is a dominant strategy in those cases as well. See \Cref{subsec:mirror_case_study} for details.

\begin{corollary}\label{cor:monomial}
Suppose $v_1(q)=q^\alpha, v_2(q)=(1-q)^\alpha$. Then for any $\alpha\geq 1$, SPA is a dominant strategy in the game and thus (SPA, SPA) is the only NE. 
\end{corollary}

\begin{figure}[ht]
\centering
\subfigure[Advertisers' value functions]{\scalebox{0.25}{\includegraphics{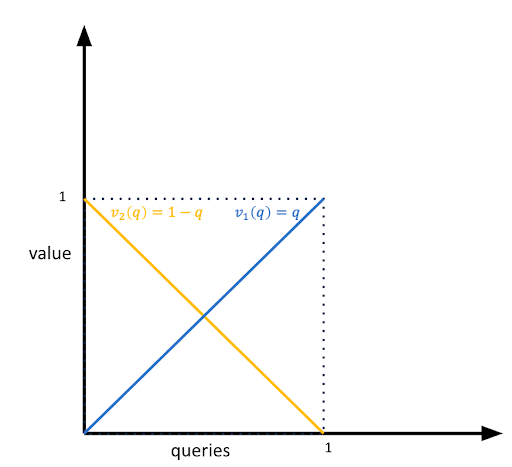}}\label{fig:sym_example_1}}
\hspace{1pt}
\subfigure[(FPA, FPA) outcome]{\scalebox{0.25}{\includegraphics{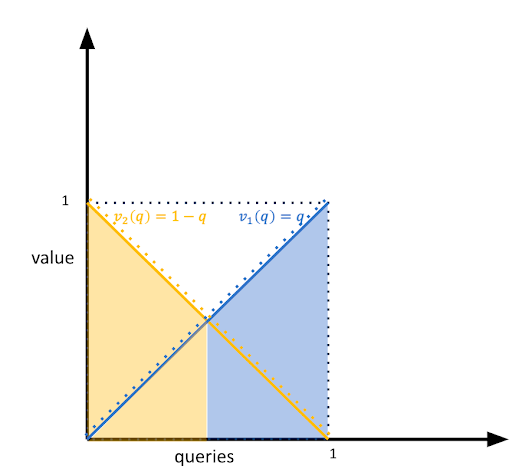}}\label{fig:sym_example_2}}
\subfigure[(SPA, SPA) outcome]{\scalebox{0.25}{\includegraphics{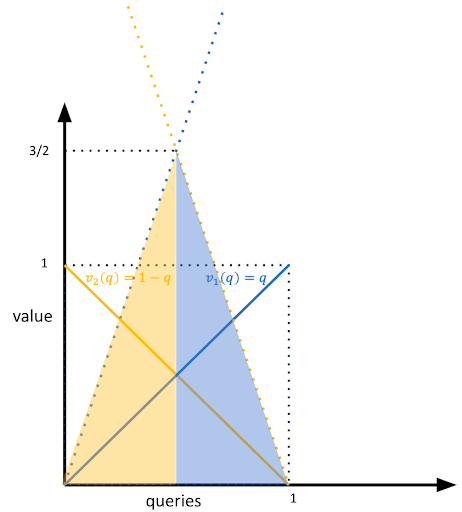}}\label{fig:sym_example_3}}
\hspace{1pt}
\subfigure[(SPA, FPA) outcome]{\scalebox{0.25}{\includegraphics{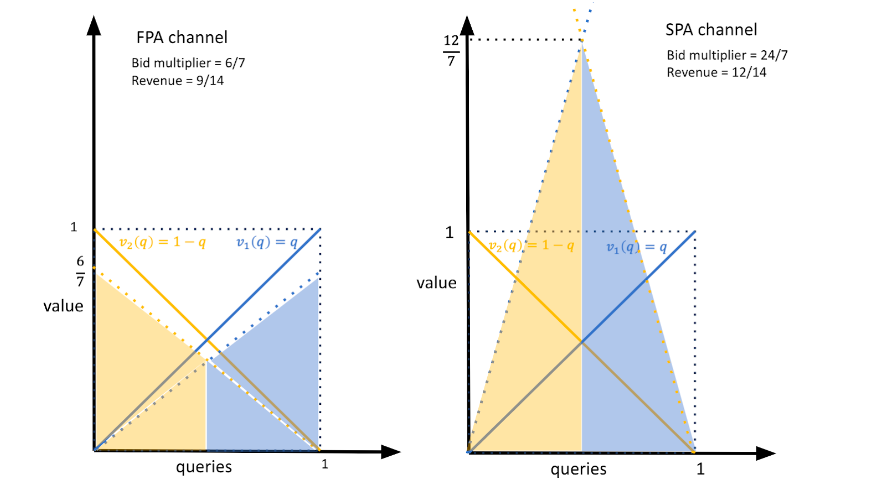}}\label{fig:sym_example_4}}
\caption{An illustration of the advertiser equilibrium in each profile.\label{fig:sym_example}}
\end{figure}

We present an illustrating example in the remainder of this section, to show the advertisers' equilibrium for each auction outcome and the platform-level game.
Suppose $v_1(q)=q, v_2(q)=1-q$ (\Cref{fig:sym_example_1}). Now $q_{\text{eff}}=\frac{1}{2}$. Figures~\ref{fig:sym_example_2}, \ref{fig:sym_example_3}, \ref{fig:sym_example_4} illustrate the outcome of the game under (FPA, FPA), (SPA, SPA) and (FPA, SPA), respectively. 

\begin{enumerate}
    \item (FPA, FPA): both advertisers use a multiplier of 1, platforms collect revenue $3/4$ (\Cref{fig:sym_example_2}).
    \item (SPA, SPA): advertiser 1 wins queries $[1/2, 1]$ which induces a total value of $3/8$. To meet the target, they set the bid multiplier such that the total payment is also $3/8$ (which implies a bid multiplier of $3$). Similarly for advertiser 2 (\Cref{fig:sym_example_3}).
    \item (SPA, FPA): we solve the problem of marginal equalization and get that both advertisers use a bid multiplier of $6/7$ in the FPA platform, and $24/7$ for the SPA platform (\Cref{fig:sym_example_4}).
\end{enumerate}

The payoff matrix for the platforms, induced by the subgame equilibria of the advertisers, is shown in Table~\ref{table:example}. We can see that (SPA, SPA) is the only NE of the game, despite the fact that revenue and allocation are identical for the (FPA, FPA) subgame.

  \begin{table}[h!]
  \caption{Induced payoff matrix for the case when $v(q) = q$.\label{table:example}}
  \centering
    \setlength{\extrarowheight}{8pt}
    \begin{tabular}{cc|c|c|}        & \multicolumn{1}{c}{} & \multicolumn{2}{c}{Platform 2}\\
      & \multicolumn{1}{c}{} & \multicolumn{1}{c}{SPA}  & \multicolumn{1}{c}{FPA} \\\cline{3-4}
      \multirow{2}*{Platform 1}  & SPA & $(3/4, 3/4)$ & $(6/7, 9/14)$ \\\cline{3-4}
      & FPA & $(9/14, 6/7)$ & $(3/4, 3/4)$ \\\cline{3-4}
    \end{tabular}
  \end{table}







\subsection{Multiple Platforms with Asymmetric Market Share}\label{subsec:sym_general}
{
In this section we generalize the result in the previous section to the setting of multiple platforms ($m\geq 2$) with identical inventory but asymmetric market share. 
For each $j\in J$, let $\gamma_j\in (0,1)$ be the market share of platform $j$. Here $\sum_j\gamma_j =1$. Advertiser $1$ has value $\gamma_j\cdot v_1(q)$ for any query in platform $j$, and advertiser $2$ has value $\gamma_j\cdot v_2(q)$.
Notice that if $\gamma_j=1$ for some $j$, then a single platform owns all the market and is indifferent between FPA and SPA, while when $m=2$ and $\gamma_1=\gamma_2=1/2$ the setting is equivalent to the one in Section~\ref{subsec:sym}.

Interestingly, we show that the same condition in \Cref{thm:symeq} holds in this general case, regardless of the number of the platforms and the market share $\{\gamma_j\}_{j\in J}$.

\begin{theorem}\label{thm:market_share_sym}
For any inefficiency-free instance of 2 advertisers and multiple platforms with asymmetric market share $\{\gamma_j\}_{j\in J}$ the following holds. If $\ela\cdot \compet > 1$, SPA is a dominant strategy and thus (SPA, SPA) is the only NE of the game. If $\ela\cdot \compet < 1$, then FPA is a dominant strategy of the game and (FPA, FPA) is the only NE. If $\ela\cdot \compet = 1$, then the game is degenerate: all profiles of the game have the same outcome. \footnote{One can easily verify that $\ela(\cdot)$ and $\compet$ are both independent of the market share $\gamma_j$.}
\end{theorem}
}


Here we give a proof sketch of \Cref{thm:market_share_sym}. Fix any platform $j$ and any auction choices of the other platforms. We prove the claim that SPA is the platform's best response if and only if $\ela\cdot \compet > 1$ and this directly implies \Cref{thm:market_share_sym}. To prove the claim, we compare the revenue of platform $j$ when it chooses SPA ($\mathcal{R}_S$) and the revenue when it chooses FPA ($\mathcal{R}_F$). As the setting is inefficiency-free, the advertisers use the same bid multiplier on any given platform. Morever, due to identical inventories across platforms, this bid multiplier is the same for all platforms that use SPA -- we denote this bid multipliers by $\mu^S(\gamma)$. Here $\gamma$ represents the total market share of the platforms choosing SPA. Similarly, the bid multiplier on all FPA platforms is the same, we denote it by $\mu^F(\gamma)$ . Observe that by merging all SPA platforms and FPA platforms respectively, the advertiser equilibrium in this multi-platform setting is equivalent to the one of the (SPA, FPA) outcome in the two-platform setting. Thus as $j$ moves from SPA to FPA, the multiplier that both advertisers use on this platform moves from $\mu^S(\gamma)$ to $\mu^F(\gamma-\gamma_j)$.\footnote{If $j$ chooses FPA, then the total market share of the SPA platforms become $\gamma-\gamma_j$.} We then derive the ratio $\frac{\mathcal{R}_S}{\mathcal{R}_F}$ as 
$$\frac{\mathcal{R}_S}{\mathcal{R}_F}=\compet\cdot \frac{\mu^S(\gamma)}{\mu^F(\gamma-\gamma_j)}=\ela\cdot \compet \cdot\frac{\mu^F(\gamma)}{\mu^F(\gamma-\gamma_j)}\quad(\text{\Cref{eq:intrabidder1}})$$
and show that $\frac{\mathcal{R}_S}{\mathcal{R}_F}>1\Leftrightarrow \ela\cdot \compet>1$ regardless of $\gamma$ and $\gamma_j$.

At a high level, the result can be interpreted as follows. When the number of platforms goes to infinity and each platform has negligible market share, the advertisers' equilibrium will (almost) not be affected by the auction choice of a single platform, or formally $\frac{\mu^F(\gamma)}{\mu^F(\gamma-\gamma_j)}$ tends to $1$. Thus SPA is the platform's best response if and only if $\ela\cdot \compet>1$. If the market share of one platform is not negligible, one can think of the platform having the option to change the auction format for part of its market share. Now similar to the above argument, the platform will have the incentive to gradually move its auction from one to another (say from FPA to SPA if $\ela\cdot \compet>1$) and will at the end chooses a unified auction format for all its market share.


\section{Efficiency vs. Bidding Reaction}
\label{sec:efficiencybidding}
In \Cref{sec:inefficiency_free} we solved the inefficiency-free setting where the allocation remains efficient for any auction change. This section tackles the case where FPA allocates more efficiently than SPA, and therefore, there is an aggregate revenue loss when a platform deviates from FPA to SPA. Now a platform faces a trade-off across efficiency loss, competition and bidding reaction in switching from FPA to SPA. While solving the model in its full generality is complex and no simple analytical solution seems to exist, we illustrate the trade-off for two particular classes of valuation functions. In the first case, we observe that when the advertisers' elasticities are small but advertiser competition $\compet$ is high, even with a small loss in efficiency, the resulting equilibrium will be (SPA, SPA). However, when the elasticities are larger and the advertiser competition $\compet$ is smaller, a small loss in efficiency can cause the equilibrium to be (FPA, FPA). 
In the second case, we observe that one advertiser's slowly decreasing elasticity and a slowly increasing advertiser competition $\compet$, paired with small efficiency losses do not suffice to exhibit multiple equilibria: the equilibrium will always be (SPA, SPA). 

\subsection{Linear vs Constant-valued Advertisers}
\label{sec:linearvsconstant}

\begin{figure}[h]
    \centering
    \includegraphics[width=0.3\textwidth]{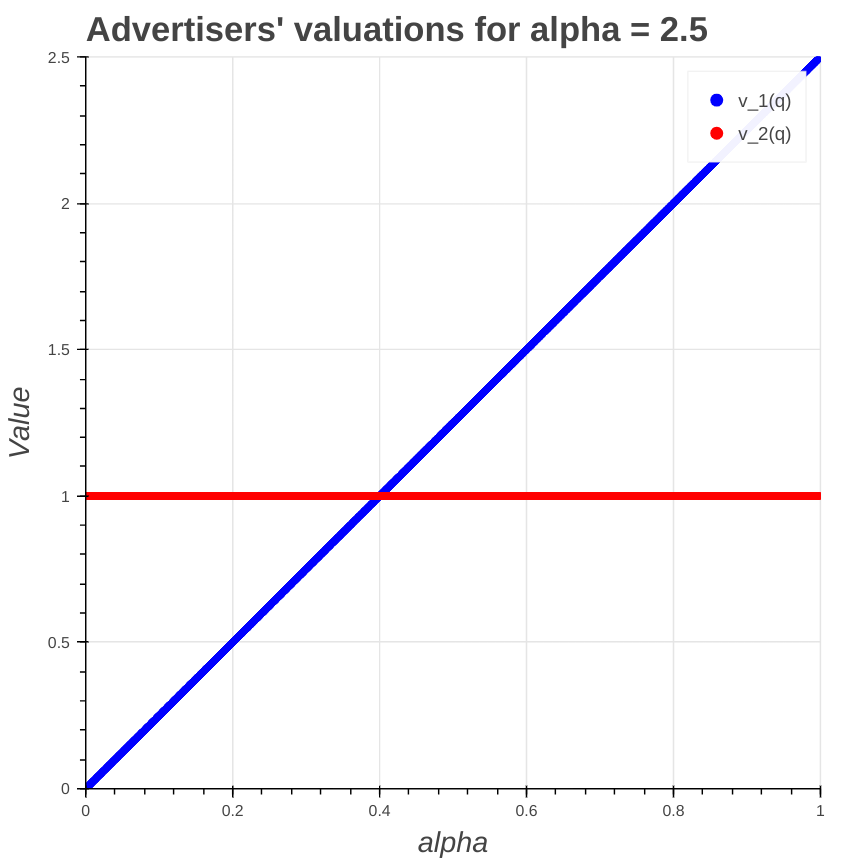}
    \caption{Plot of valuations $v_1(q) = \alpha q$ (for $\alpha = 2.5$) and $v_2(q) = 1$.}
    \label{fig:linearExample}
\end{figure}

We first consider two symmetric platforms and two advertisers, one with valuation $v_{1}(q) = \alpha q$ (for some $\alpha \in (2, 4)$) and $v_2(q) = 1$ (see Figure~\ref{fig:linearExample} for an example). We restrict our attention to the case where $\alpha \in (2,4)$ because for values outside this range, the equilibria of the game result in non-interior solutions. 

\begin{claim}
\label{cl:alpharange}
For $\alpha \not \in (2,4)$, in any advertisers' equilibrium of the (SPA, SPA) auction profile, one advertiser will win all queries in both platforms. 
\end{claim} 

It will follow from our analysis that advertiser 2's optimal bid multiplier in the second price platform is $2$, which is independent of $\alpha$. This means that, no matter what, advertiser $2$ will bid $2$ on every query sold in a second price auction. If $\alpha \leq 2$, advertiser $2$ will outright win all queries in this platform since it would be prohibitive for advertiser $1$ to bid above $2$. On the other hand, if $\alpha \geq 4$, then the total value of advertiser 1 for all queries is greater than 2, which is the total value of the bids placed by advertiser 2. Therefore, advertiser 1 would be fine submitting an arbitrarily high bid on the second price platform and winning (virtually) all queries. In fact, for any bid multiplier $M$ chosen by advertiser 1, a bid multiplier $M' > M$ will be strictly preferable. This implies that for $\alpha \geq 4$ there is no equilibrium. Thus $\alpha \in (2,4)$ is the sweet-spot for this family of valuations where advertiser 1's value is high enough to win some queries by shading their bid lightly, but not high enough to be able to outright win all of them. 


We now state the main result of this section. 

\begin{theorem}\label{thm:mainconstantvslinear}
The platforms' equilibria of the game in the linear-vs-constant setting depend on the value of $\alpha$ as follows. There exist $\alpha_1 \leq \alpha_2 \leq \alpha_3 \leq \alpha_4$ s.t.:
\begin{itemize}
    \item for $\alpha \in (2, \alpha_1)$ there are three Nash equilibria: (SPA, FPA), (FPA, SPA) and a mixture of both, 
    \item for $\alpha \in [\alpha_1, \alpha_2)$, there is a unique Nash equilibrium: (SPA, SPA),
    \item for $\alpha \in [\alpha_2, \alpha_3)$, there are three Nash equilibria: (FPA, FPA), (SPA, SPA) and a mixture of both,
    \item for $\alpha \in [\alpha_3, 4)$, there is a unique Nash equilibrium: (FPA, FPA).
\end{itemize}
\end{theorem}

\begin{figure}[h!]
\centering
        \includegraphics[width=0.3\textwidth]{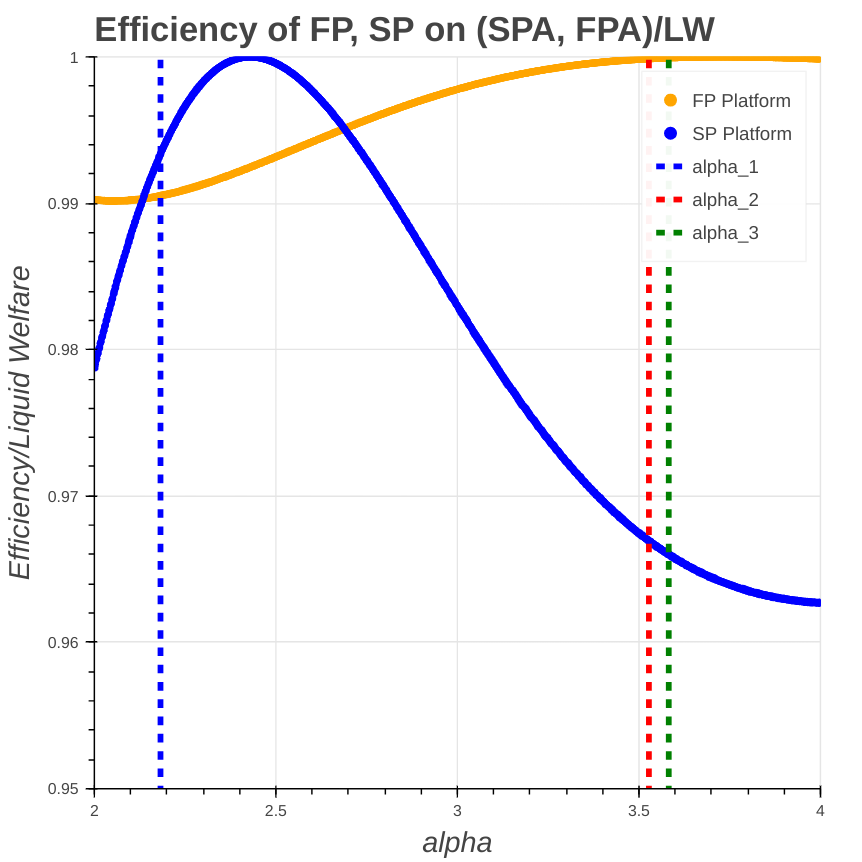}
    \caption{This plot shows the efficiency of the FP and SP platforms on the (SPA, FPA) subgame.}
    \label{fig:SPFPeff}
\end{figure}

\begin{figure}
    \centering
        \includegraphics[width=0.3\textwidth]{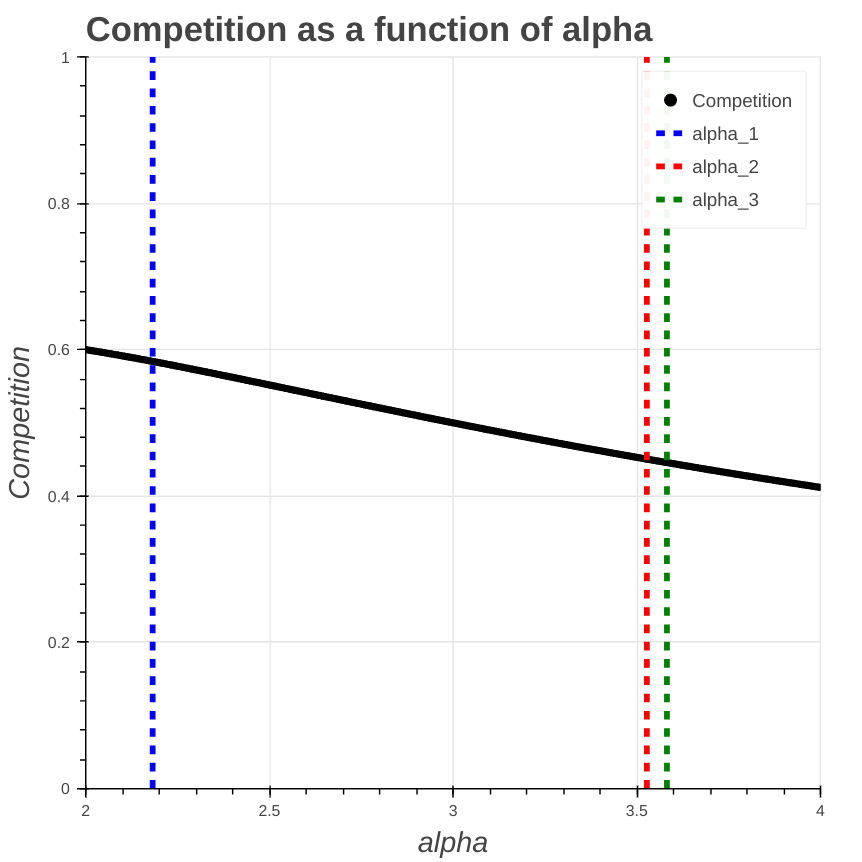}
    \caption{This plot shows the competition metric $\compet$ as a function $\alpha$ for the case of a linear and constant-valued advertiser.}
    \label{fig:competition}
\end{figure}

\begin{figure}
\centering
\includegraphics[width=0.3\textwidth]{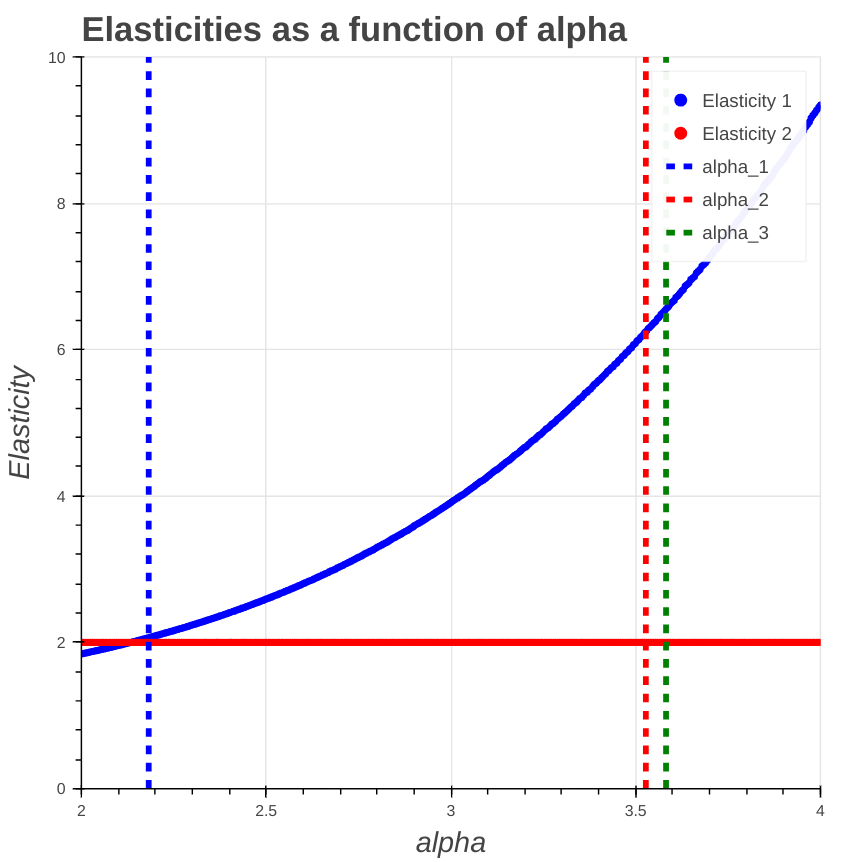}
        \caption{This plot shows the elasticity of both advertisers as a function $\alpha$ for the case of a linear and constant-valued advertiser.}
    \label{fig:elasticity}
\end{figure}

Without getting in to the proof details of Theorem~\ref{thm:mainconstantvslinear} (whose proof is deferred to Appendix~\ref{app:linearvsconstant}), it is worth observing the plots of efficiency, competition and elasticities as a function of the parameter $\alpha$ (see Figures~\ref{fig:SPFPeff},~\ref{fig:competition},~\ref{fig:elasticity}). 
As $\alpha$ increases, there is a sharp loss of efficiency on the SP platform relative to the FP platform. This, combined with a decrease in advertiser competition $\compet$, is sufficient to overcome an increase in elasticity and shift the equilibrium from (SPA, SPA) to (FPA, FPA) (for high values of $\alpha$). This observation contrasts with the results from Section~\ref{sec:inefficiency_free} where advertiser competition and elasticity were the only driving factors in determining equilibria. Here we observe that efficiency losses may also play a substantial role in determining equilibria.    

[h!]
[h!]

\subsection{Exponential Valuations}
\label{sec:expvsexp}

We next consider the case where both advertisers have exponential valuations $v_{1}(q) = \alpha e^{-q}$, $v_{2}(q) = e^{-2q}$. In this case, we extend the continuum of queries to the entire real line (i.e., $q \in [0, \infty)$). We also restrict our attention to a particular subset of values for $\alpha$, as values outside of this range would yield non-interior solutions. 

\begin{figure}[h!]
    \centering
        \includegraphics[width=0.3\textwidth]{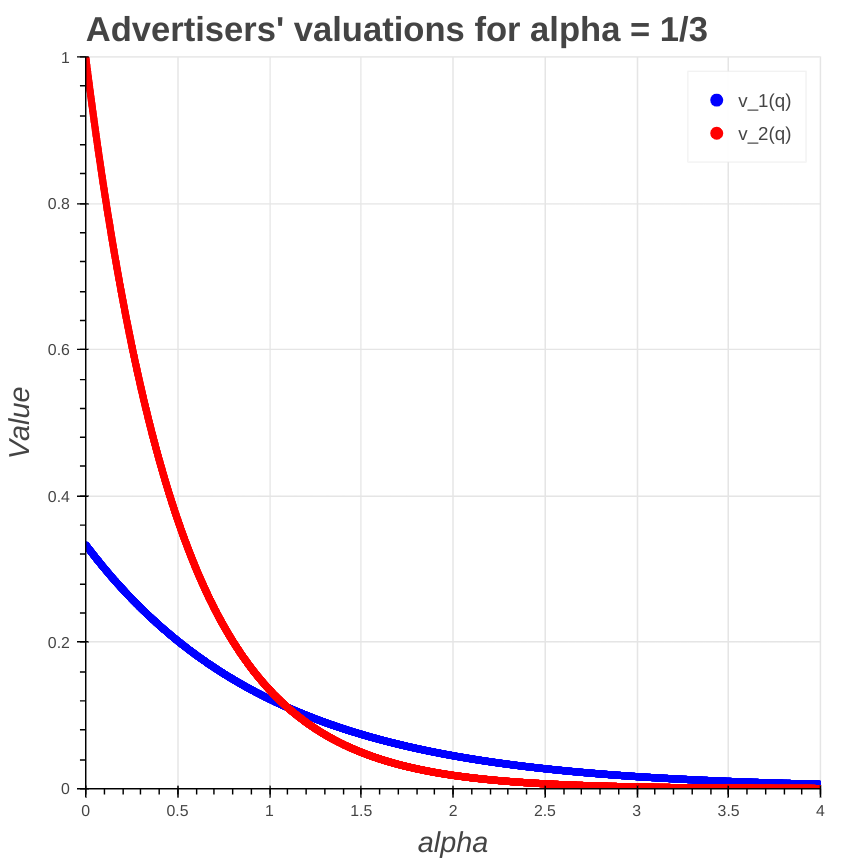}
    \caption{This plot shows the both advertisers valuations for $\alpha = 1/3$.}
    \label{fig:valuations_exp}
\end{figure}
\begin{figure}[h!]
    \centering
        \includegraphics[width=0.3\textwidth]{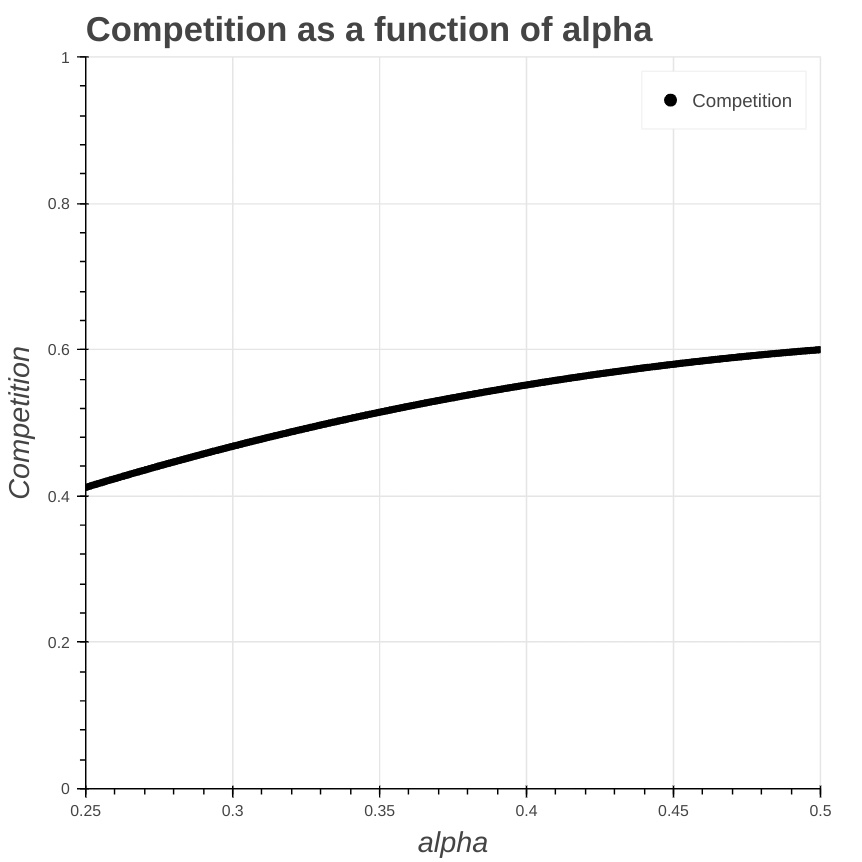}
    \caption{This plot shows the competition metric $\compet$ as a function $\alpha$ for the case of two exponential valued advertisers.}
    \label{fig:competition_exp}
\end{figure}
\begin{figure}[h!]
\centering
\includegraphics[width=0.3\textwidth]{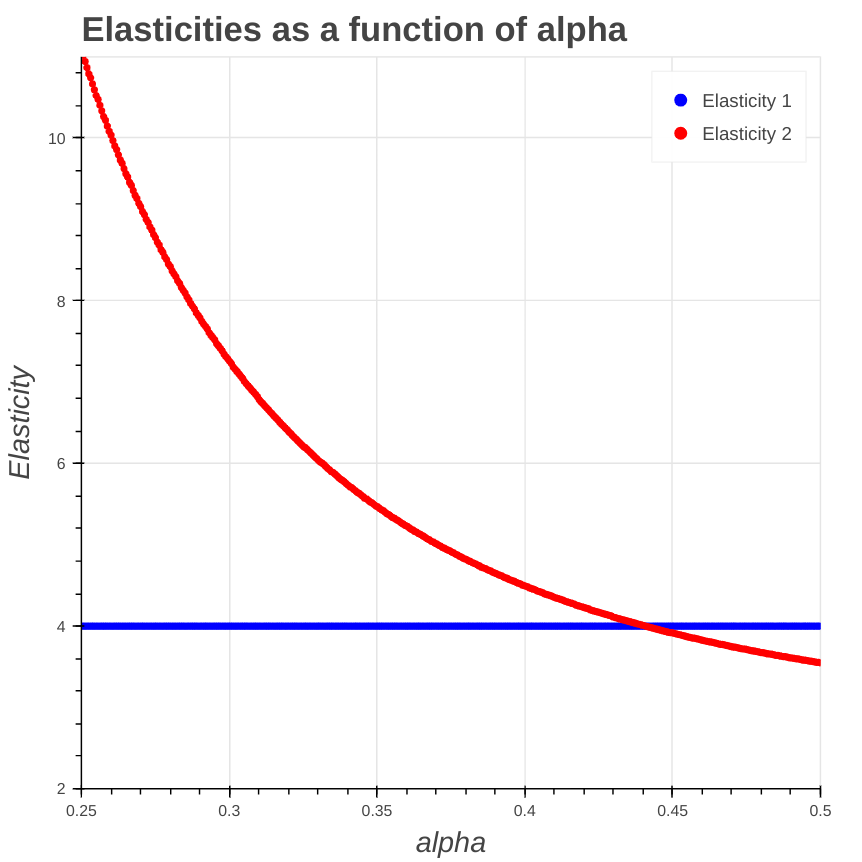}
        \caption{This plot shows the elasticity of both advertisers as a function $\alpha$ for the case of a two exponential-valued advertisers.}
    \label{fig:elasticity_exp}
\end{figure}

\begin{claim}
\label{cl:expalpha}
For this case, when $\alpha \not \in (1/4, 1/2)$, either a single advertiser wins all queries or there is no equilibrium in the (SPA, SPA) subgame. 
\end{claim}

The proof of Claim~\ref{cl:expalpha} will be apparent from the analysis of the (SPA, SPA) subgame, so we do not provide a separate proof for it. We now state the main result of this subsection. 

\begin{theorem}\label{thm:expequilibria}
For any $\alpha\in (\frac{1}{4}, \frac{1}{2})$, SPA is a dominant strategy. Hence (SPA, SPA) is the only NE of the game.
\end{theorem}

In contrast to the results from Section~\ref{sec:inefficiency_free} and Section~\ref{sec:linearvsconstant}, in this case we observe a small decrease in one of the advertisers' elasticities and a steeper increase in competition as $\alpha$ increases. The loss of efficiency in this case does not suffice to allow for multiple equilibria. The proof of Theorem~\ref{thm:expequilibria} can be found in Appendix~\ref{app:expvsexp}.


\newpage
\bibliographystyle{plainnat}
\bibliography{bibliography}

\begin{thebibliography}{17}
\providecommand{\natexlab}[1]{#1}
\providecommand{\url}[1]{\texttt{#1}}
\expandafter\ifx\csname urlstyle\endcsname\relax
  \providecommand{\doi}[1]{doi: #1}\else
  \providecommand{\doi}{doi: \begingroup \urlstyle{rm}\Url}\fi

\bibitem[Agarwal et~al.(2014)Agarwal, Ghosh, Wei, and
  You]{10.1145/2623330.2623366}
Deepak Agarwal, Souvik Ghosh, Kai Wei, and Siyu You.
\newblock Budget pacing for targeted online advertisements at linkedin.
\newblock In \emph{Proceedings of the 20th ACM SIGKDD International Conference
  on Knowledge Discovery and Data Mining}, KDD '14, page 1613–1619, New York,
  NY, USA, 2014. Association for Computing Machinery.
\newblock ISBN 9781450329569.
\newblock \doi{10.1145/2623330.2623366}.
\newblock URL \url{https://doi.org/10.1145/2623330.2623366}.

\bibitem[Aggarwal et~al.(2019)Aggarwal, Badanidiyuru, and
  Mehta]{aggarwal_ashwin_mehta_www19}
Gagan Aggarwal, Ashwinkumar Badanidiyuru, and Aranyak Mehta.
\newblock Autobidding with constraints.
\newblock In \emph{Web and Internet Economics: 15th International Conference,
  WINE 2019, New York, NY, USA, December 10--12, 2019, Proceedings 15}, pages
  17--30. Springer, 2019.

\bibitem[Aggarwal et~al.(2023)Aggarwal, Perlroth, and
  Zhao]{aggarwal_perlroth_zhao_ec23}
Gagan Aggarwal, Andres Perlroth, and Junyao Zhao.
\newblock Multi-channel auction design in the autobidding world.
\newblock In \emph{Proceedings of the 24th ACM Conference on Economics and
  Computation}, EC '23, page~21, New York, NY, USA, 2023. Association for
  Computing Machinery.
\newblock ISBN 9798400701047.
\newblock \doi{10.1145/3580507.3597707}.
\newblock URL \url{https://doi.org/10.1145/3580507.3597707}.

\bibitem[Akbarpour and Li(2020)]{credibility_2020}
Mohammad Akbarpour and Shengwu Li.
\newblock Credible auctions: A trilemma.
\newblock \emph{Econometrica}, 88\penalty0 (2):\penalty0 425--467, 2020.
\newblock \doi{https://doi.org/10.3982/ECTA15925}.
\newblock URL \url{https://onlinelibrary.wiley.com/doi/abs/10.3982/ECTA15925}.

\bibitem[Alimohammadi et~al.(2023)Alimohammadi, Mehta, and
  Perlroth]{yeganeh_mehta_perlroth_ec23}
Yeganeh Alimohammadi, Aranyak Mehta, and Andres Perlroth.
\newblock Incentive compatibility in the auto-bidding world.
\newblock In \emph{Proceedings of the 24th ACM Conference on Economics and
  Computation}, EC '23, page~63, New York, NY, USA, 2023. Association for
  Computing Machinery.
\newblock ISBN 9798400701047.
\newblock \doi{10.1145/3580507.3597725}.
\newblock URL \url{https://doi.org/10.1145/3580507.3597725}.

\bibitem[Balseiro and Gur(2019)]{balseiro_gur_2019}
Santiago~R. Balseiro and Yonatan Gur.
\newblock Learning in repeated auctions with budgets: Regret minimization and
  equilibrium.
\newblock \emph{Management Science}, 65\penalty0 (9):\penalty0 3952--3968,
  2019.
\newblock \doi{10.1287/mnsc.2018.3174}.
\newblock URL \url{https://doi.org/10.1287/mnsc.2018.3174}.

\bibitem[Bateni et~al.(2014)Bateni, Feldman, Mirrokni, and Wong]{batenietal14}
MohammadHossein Bateni, Jon Feldman, Vahab Mirrokni, and Sam Chiu-wai Wong.
\newblock Multiplicative bidding in online advertising.
\newblock In \emph{Proceedings of the Fifteenth ACM Conference on Economics and
  Computation}, EC '14, page 715–732, New York, NY, USA, 2014. Association
  for Computing Machinery.
\newblock ISBN 9781450325653.
\newblock \doi{10.1145/2600057.2602874}.
\newblock URL \url{https://doi.org/10.1145/2600057.2602874}.

\bibitem[Boyd and Vandenberghe(2004)]{boyd2004convex}
Stephen~P Boyd and Lieven Vandenberghe.
\newblock \emph{Convex optimization}.
\newblock Cambridge university press, 2004.

\bibitem[Conitzer et~al.(2022{\natexlab{a}})Conitzer, Kroer, Panigrahi,
  Schrijvers, Stier-Moses, Sodomka, and Wilkens]{doi:10.1287/mnsc.2022.4310}
Vincent Conitzer, Christian Kroer, Debmalya Panigrahi, Okke Schrijvers,
  Nicolas~E. Stier-Moses, Eric Sodomka, and Christopher~A. Wilkens.
\newblock Pacing equilibrium in first price auction markets.
\newblock \emph{Management Science}, 68\penalty0 (12):\penalty0 8515--8535,
  2022{\natexlab{a}}.
\newblock \doi{10.1287/mnsc.2022.4310}.
\newblock URL \url{https://doi.org/10.1287/mnsc.2022.4310}.

\bibitem[Conitzer et~al.(2022{\natexlab{b}})Conitzer, Kroer, Sodomka, and
  Stier-Moses]{10.1287/opre.2021.2167}
Vincent Conitzer, Christian Kroer, Eric Sodomka, and Nicolas~E. Stier-Moses.
\newblock Multiplicative pacing equilibria in auction markets.
\newblock \emph{Oper. Res.}, 70\penalty0 (2):\penalty0 963–989, mar
  2022{\natexlab{b}}.
\newblock ISSN 0030-364X.
\newblock \doi{10.1287/opre.2021.2167}.
\newblock URL \url{https://doi.org/10.1287/opre.2021.2167}.

\bibitem[Deng et~al.(2021)Deng, Mao, Mirrokni, and Zuo]{yuan_towards}
Yuan Deng, Jieming Mao, Vahab Mirrokni, and Song Zuo.
\newblock Towards efficient auctions in an auto-bidding world.
\newblock In \emph{Proceedings of the Web Conference 2021}, WWW '21, page
  3965–3973, New York, NY, USA, 2021. Association for Computing Machinery.
\newblock ISBN 9781450383127.
\newblock \doi{10.1145/3442381.3450052}.
\newblock URL \url{https://doi.org/10.1145/3442381.3450052}.

\bibitem[Deng et~al.(2023)Deng, Golrezaei, Jaillet, Liang, and
  Mirrokni]{deng2023multi}
Yuan Deng, Negin Golrezaei, Patrick Jaillet, Jason Cheuk~Nam Liang, and Vahab
  Mirrokni.
\newblock Multi-channel autobidding with budget and roi constraints.
\newblock In \emph{Proceedings of the 40th International Conference on Machine
  Learning}, ICML'23. JMLR.org, 2023.

\bibitem[Despotakis et~al.(2021)Despotakis, Ravi, and Sayedi]{marketing2021}
Stylianos Despotakis, R.~Ravi, and Amin Sayedi.
\newblock First-price auctions in online display advertising.
\newblock \emph{Journal of Marketing Research}, 58\penalty0 (5):\penalty0
  888--907, 2021.
\newblock \doi{10.1177/00222437211030201}.
\newblock URL \url{https://doi.org/10.1177/00222437211030201}.

\bibitem[Google(2021)]{google21}
Google.
\newblock Moving adsense to a first-price auction, 2021.
\newblock URL
  \url{https://blog.google/products/adsense/our-move-to-a-first-price-auction/}.

\bibitem[Paes~Leme et~al.(2020)Paes~Leme, Sivan, and
  Teng]{renato_balu_yifeng_www2020}
Renato Paes~Leme, Balasubramanian Sivan, and Yifeng Teng.
\newblock Why do competitive markets converge to first-price auctions?
\newblock In \emph{Proceedings of The Web Conference 2020}, WWW '20, page
  596–605, New York, NY, USA, 2020. Association for Computing Machinery.
\newblock ISBN 9781450370233.
\newblock \doi{10.1145/3366423.3380142}.
\newblock URL \url{https://doi.org/10.1145/3366423.3380142}.

\bibitem[Susan et~al.(2023)Susan, Golrezaei, and Schrijvers]{susan2023multi}
Fransisca Susan, Negin Golrezaei, and Okke Schrijvers.
\newblock Multi-platform budget management in ad markets with non-ic auctions.
\newblock \emph{arXiv preprint arXiv:2306.07352}, 2023.

\bibitem[Xu et~al.(2015)Xu, Lee, Li, Qi, and Lu]{10.1145/2783258.2788615}
Jian Xu, Kuang-chih Lee, Wentong Li, Hang Qi, and Quan Lu.
\newblock Smart pacing for effective online ad campaign optimization.
\newblock In \emph{Proceedings of the 21th ACM SIGKDD International Conference
  on Knowledge Discovery and Data Mining}, KDD '15, page 2217–2226, New York,
  NY, USA, 2015. Association for Computing Machinery.
\newblock ISBN 9781450336642.
\newblock \doi{10.1145/2783258.2788615}.
\newblock URL \url{https://doi.org/10.1145/2783258.2788615}.

\end{thebibliography}

\appendix
\section*{Appendix}

\section{Missing Proofs from Section~\ref{sec:advertiser_subgame}}\label{app:sec_optimal_bidding}

\begin{proof}[Proof of \Cref{th1:opt_bidding}]
To show the existence of a solution, observe that bidding $\mu_j=0$ for all $j\in J$ is a feasible solution and therefore the optimization set is not empty. Moreover, because $V_j(\hm_j)<\infty$ we have that writing Problem~\eqref{bidder-problem} as supremum is well-defined.

Observe that if for all $j$ we have that $\hm_j < \infty$ then we can restrict to a compact set and because $V_j(\mu_j)$ are continuous a solution exists. 

On the other hand, if for some $\hat j$, $\hm_{\hat j} = \infty$, then let $(\mu^n_j)_{j\in J}$ be a sequence of feasible bid multipliers approaching the supremum of Problem~\eqref{bidder-problem}. We assert that $\mu^n_{\hat j} < \infty$. Suppose for the sake of a contradiction that this is not the case and $\mu^n_{\hat j} \to \infty$. 
\begin{itemize}
    \item If $\hat j\in J_F$, then because $V_j$ are increasing and bounded, we get that
$$     \frac{\sum_{j\in J} \mu^n_j V_j(\mu^n_j) - \sum_{j\in J_S}\int_0^{\mu^n_j} V_j(z)dz } { \sum_{j\in J} V_j(\mu^n_j) } >\frac{\mu^n_{\hat j} V_{\hat j}(\mu^n_{\hat j})}{\sum_{j\in J} V(\hm_j)} .$$
The right hand side of the inequality goes to infinity as $n\to \infty$, which implies that $(\mu^n_j)_{j\in J}$ does not satisfy the target constraint (Equation~\eqref{eq:target-constraint}) for large $n$. This is a contradiction.
\item If $\hat j \in J_S$, from Condition~\eqref{eq:condtion_hm_infty} we must have that in one of the platforms $\tilde j$ the advertiser is bidding less than $\hm_{\tilde j}$ to make the target constraint feasible. Hence, for all $n$ we have that $\mu^n_{\tilde j} \leq M< {\hm}_{\tilde j}$ for some constant $M$. Using Lemma~\ref{lem0:marginals} we obtain that $MC_{\hat j}(\mu^n_{\hat j})\geq \mu^n_{\hat j} > C > MC_{\tilde j}(\mu^n_{\tilde j}) $ for some constant $C$. Because $\mu^n_{\tilde j}$ is bounded, we can assume that it converges and, therefore, we can take $\epsilon>0$ such that for $n$ large enough if we replace the bids $\mu^n_{\hat j}$ by $\mu^n_{\hat j} - \Delta_{\hat j}$ and $\mu^n_{\tilde j}$ by $\mu^n_{\tilde j} +  \Delta_{\tilde j}$, the new vector of bids is feasible and increases the total value for the advertiser by $\epsilon$. Because $\epsilon$ is independent of $n$ this contradicts the assumption that the original sequence $(\mu_j^n)_{j\in J}$ is approaching the supremum.
\end{itemize}
We therefore conclude that if for some $\hat j$, $\hm_{\hat j} = \infty$ then any sequence approximating the supremum lies on a bounded set. Because $(V_j)_{j\in J}$ are continuous functions we have that a solution exists.
\end{proof}

\begin{proof}[Proof of \Cref{lem0:marginals}]
Because $V_j$ is twice differentiable on $[0,\hm_j]$, we have from L'Hopital's rule that 
$$MC_j(\mu_j) = \frac{C_j'(\mu_j)}{V_j'(\mu_j)}.$$
Then, the characterization of $MC_j$ immediately follows by noticing that if $j\in J_S$ then $C'_j(\mu_j) = \mu_jV'_j(\mu_j)$, and if $j\in J_F$ then $C'_j(\mu_j) = \mu_jV'_j(\mu_j) + V_j(\mu_j)$. 

As for the monotonicity of $MC_j$, if $j\in J_S$, $MC_j$ is increasing. For $j\in J_F$, since $V_j$ is twice differentiable, it suffices to show that $MC'_j(\mu_j)>0$ for all $\mu_j$. 
Indeed, observe that
$$ MC'_j(\mu_j) = 2 - \frac{V_j(\mu_j)V_j''(\mu_j)}{{V_j'(\mu_j)}^2}.$$
We conclude that  $MC_j(\mu_j)'>0$ since $V_j$ is concave, and therefore, $V_j''(\mu_j) \leq 0$. 
\end{proof}

\begin{proof}[Proof of \Cref{theo:marg_sufficient_condition}]
We split the proof in two steps: Step 1 shows the uniqueness of the solution to~\eqref{eq:marg_equalization}. Step 2 shows the existence of a solution. Step 3 shows the uniqueness of the solution to Problem~\eqref{bidder-problem}. \medskip

\noindent{\bf Step 1.} First, notice that if $(\mu^*_j)_{j\in J}$ is a nonzero solution to~\eqref{eq:marg_equalization} we must have that at least for one of the platforms, say $j'$, the advertiser uses a bid multiplier $\mu^*_{j'}\geq 1$. Otherwise, if all bid multipliers are less than $1$, in each of the platforms the average cost would be less than $1$ which contradicts that the target constraint (Equation~\eqref{eq:target-constraint}) holds with equality. Hence, from Lemma~\ref{lem0:marginals} and the fact the marginal cost are the same in each of the platforms, we conclude that for all $j\in J$, $MC_j(\mu^*_j) = MC_{j'}(\mu^*_{j'}) \geq \mu^*_{j'}\geq   1$.

To show the uniqueness of the solution to~\eqref{eq:marg_equalization}, consider for the sake of contradiction, two different nonzero solutions $(\mu^1_j)_{j\in J}$ and $(\mu^2_j)_{j\in J}$ with $\mu^1_{j'} < \mu^2_{j'}$ for some platform $j'$. Therefore, for $j \in J$ we obtain that
$$MC_{j}(\mu^1_{j}) = MC_{j'}(\mu^1_{j'}) < MC_{j'}(\mu^2_{j'}) = MC_j(\mu^2_j),$$
where the first and third equality hold because both bid multipliers are solving Equation~\eqref{eq:marg_equalization}, the second strict inequality holds because marginal cost is a strictly increasing function. Using that the marginal cost functions are strict we conclude that $\mu^1_{j} < \mu^2_{j}$ for all $j\in J$. 

To conclude the proof of uniqueness, observe that
$$C_j(\mu^2_j)- C_j(\mu^1_j) = \int_{\mu^1_j}^{\mu^2_j} MC_j(z)V'_j(z)dz >  \int_{\mu^1_j}^{\mu^2_j} V'_j(z)dz = V_j(\mu^2_j)- V_j(\mu^1_j), $$
where the strict inequality comes from monotonicity of the marginal cost functions and the result of the first paragraph so that $MC_j(z)>MC_j(\mu^1_j)\geq 1$. Therefore,
\begin{align*}
0 & = \sum_{j\in J} V_j(\mu^1_j)- C_j(\mu^1_j) \\
&=  \sum_{j\in J} V_j(\mu^2_j) - C_j(\mu^2_j) - \underbrace{\left (V_j(\mu^2_j)- V_j(\mu^1_j)) -  (C_j(\mu^2_j)- C_j(\mu^1_j)) \right)}_{>0} \\
&< \sum_{j\in J} V_j(\mu^2_j) - C_j(\mu^2_j) = 0
\end{align*}
where the first and last equality hold because both bid multipliers solve~\eqref{eq:marg_equalization}. This is a contradiction. Therefore there is a unique solution to the system. \medskip

\noindent{\bf Step 2.} To show the existence of a solution we assert that if for some platform $\hat j \in J_S$ we have that $\hat \mu_{\hat j} = \infty$, then condition~\eqref{eq:condtion_hm_infty} holds. Indeed consider $(\mu_j)_{j\in J}$ the non trivial solution to Equation~\eqref{eq:marg_equalization}. Then $\mu_{\hat j}<\hat \mu_{\hat j}$ and $\mu_j \leq \hm_{j}$ for all $j\in J$. Therefore, from the logic of the Step 1 replacing $(\mu^1_j)$ by the solution $(\mu_j)_{j\in J}$ and $(\mu^2_j)$ by $(\hm_j)_{j\in J}$ we conclude that the target-constraint (Equation~\eqref{eq:target-constraint} at $(\hm_j)_{j\in J}$ is not feasible. Thus, from \Cref{th1:opt_bidding} we conclude that a solution to Problem ~\eqref{bidder-problem} exists. \medskip

\noindent{\bf Step 3.} Consider a solution $(\mu_j)_{j\in J}$ of Problem ~\eqref{bidder-problem}, which exists due to Theorem~\ref{th1:opt_bidding}. Using the Karush-Kuhn-Tucker Theorem (see \citet{boyd2004convex} for a textbook treatment) consider the Lagrangian of Problem ~\eqref{bidder-problem} as $$\mathcal{L}(\bm{\mu}, \lambda) = \sum_{j \in J} V_j(\mu_j)(1 + \lambda(\mu_j-1)) - \lambda \sum_{j\in J_S} \int_0^{\mu_j} V_j(z)dz,$$
where $\bm{\mu}\geq 0$ is the vector of bid multipliers and $\lambda\geq 0$ is the Lagrangian dual variable. Then at $(\mu_j)_{j\in J}$ we have that for the dual solution $\lambda\geq 0 $,
$$
\frac{\partial_j}{\partial \mu_j} \mathcal{L}(\bm{\mu}, \lambda)
\begin{cases}
\leq 0 &\mbox{ if } \mu_j = 0 \\
= \,0 &\mbox{ if } \mu_j \in (0,\hm_j) \\
\geq 0 &\mbox{ if } \mu_j = \hm_j 
\end{cases}.
$$
We assert that $\lambda >0$. Indeed if $\lambda = 0$, then $\frac{\partial_j}{\partial \mu_j} \mathcal{L}(\bm{\mu}, \lambda) = V'_j(\mu_j)$ and because $V_j$ is an increasing function this would imply that for all $j$, $\mu_j = \hm_j$. From Step 2 we know that this is not feasible. Thus, $\lambda>0$. 

Because $\lambda>0$, the KKT conditions can be rewritten as 
$$
\begin{cases}
    MC_j(\mu_j) \geq \frac{\lambda - 1}{\lambda} &\mbox{ if } \mu_j = 0 \\
     MC_j(\mu_j) = \frac{\lambda - 1}{\lambda} &\mbox{ if } \mu_j \in (0,\hm_j) \\
    MC_j(\mu_j) \leq \frac{\lambda - 1}{\lambda} &\mbox{ if } \mu_j = \hm_j 
\end{cases}.$$
Notice that since there is a non-zero solution to Problem~\eqref{bidder-problem} (take the feasible solution $(\mu^*_j)_{j\in J}$ of~\eqref{eq:marg_equalization}), we have that $\lambda>1$. This implies that $\mu_j>0$ for all $j$. 

To conclude we separate the analysis into two scenarios. In the first scenario, if an optimal bidding solution is such that $\mu_j<\hm_j$, then we have the solution solves~\eqref{eq:marg_equalization}. Hence, from Step 1, the solution has to be $(\mu^*_j)_{j\in J}$.
For the second scenario, if the optimal solution $(\mu_j)_{j\in J}$ is such that $\mu_{j'}=\hm_{j'}$ for some platform $j'$. Using that $\mu^*_{j'} < \hm_{j'} = \mu_{j'}$, we derive that
$$ MC_j(\mu^*_j)  = MC_{j'}(\mu^*_{j'}) <  MC_{j'}(\mu_{j'}) \leq  MC_{j}(\mu_{j}), $$
where the first equality holds because $\mu^* $ solves the system ~\eqref{eq:marg_equalization}, the strict inequality comes because $MC_{j}$ is increasing, the last equality comes from the KKT conditions. 
From the last chain of inequalities, we can replicate the same argument as in Step 1 by interchanging $(\mu^*_j)_{j\in J}$ by $(\mu^1_j)_{j\in J}$ and $(\mu_j)_{j\in J}$ by $(\mu^2_j)_{j\in J}$ and we get a contradiction. Therefore, there is a unique solution to Problem~\eqref{bidder-problem}.
\end{proof}

\begin{proof}[Proof of Theorem~\ref{theo:sufficient_condition_general_statement}]
First, notice that using a bid multiplier $\mu_j=1$ for all $j\in J$ is a feasible solution. This implies that $(\mu^*_j)_{j\in J} $ cannot be the zero for all $j$.

Second, deriving the Karush-Kuhn-Tucker conditions that have to hold at $(\mu^*_j)_{j\in J} $ as in Step 2 of the Proof of Theorem~\ref{theo:marg_sufficient_condition}, we conclude that $(\mu^*_j)_{j\in J} $ is a non-zero solution to~\eqref{eq:marg_equalization}. We conclude the proof from Theorem~\ref{theo:marg_sufficient_condition}.  
\end{proof}

\begin{proof}[Proof of \Cref{coro:vj_conditions}]
From \Cref{lem0:marginals} we have that the marginal costs functions are increasing on $[0,\hm_j]$. Because there is a platform $\hat j\in J_F$ with $\hm_{\hat j} = \infty$, then for any sequence of bid multipliers $(\mu^n_j)$ with $\mu^n_j \to \hm_j$, we have that
$$ \frac{\sum_{j\in J} \mu^n_j V_j(\mu^n_j) - \sum_{j\in J_S}\int_0^{\mu^n_j} V_j(z)dz } { \sum_{j\in J} V_j(\mu^n_j) } >\frac{\mu^n_{\hat j} V_{\hat j}(\mu^n_{\hat j})}{\sum_{j\in J} V(\hm_j)} $$
the right hand side goes to infinity as $n\to \infty$. Therefore, condition~\eqref{eq:condtion_hm_infty} holds and from \Cref{th1:opt_bidding} we have that a solution to Problem~\eqref{bidder-problem}, $(\mu^*_j)_{j\in J} $ exists. Therefore, $\mu^*_j<\infty = \hm_j$. All the conditions for \Cref{theo:sufficient_condition_general_statement} hold which allow us to conclude.
\end{proof}

\begin{proof}[Proof of \Cref{theo:fpa_subgame}]
From the symmetry refinement (R3) (see \Cref{as:refinments}), we have that both bidders have to bid the same in each platform. Therefore, the problem is equivalent to solving the FPA game in a single platform. Theorem 6.5 in \citet{yuan_towards} shows that the unique equilibrium with undominated strategies is to bid using a bid multiplier equal to the advertiser's target CPA. In our case, this means $\mu^*_i = 1$. The authors further show that the equilibrium is efficient so that and the platform gets revenue equal to the optimal liquid welfare. In our two platform case this means that the equilibrium query-threshold is $q_{\text{eff}}$ and each platform obtains $W^*/2$. 
\end{proof}

\begin{proof}[Proof of \Cref{theo:spa_subgame}]
From the symmetry refinement (R3) (see \Cref{as:refinments}), we have that both bidders have to bid the same in each platforms. Therefore, the problem is equivalent to solving the SPA game in a single platform.
The proof of existence and uniqueness for a single platform can be found in Theorem 5.7 of \citet{yeganeh_mehta_perlroth_ec23}, where given those conditions the equilibrium of SPA is unique which also implies that it is incentive compatible for autobidders. 
\end{proof}

\begin{proof}[Proof of \Cref{lem:subgame_fpa_spa}]
Fix a bid multiplier $\mu_1>0$. First, observe that for any bid multiplier $\mu$ that advertiser 2 use the query-threshold is $q(\mu_1,\mu) = h^{-1}(\frac \mu {\mu_1})$ for $\hm_j = h(0)\mu_1$. 
To show that $V_{2j}$ is increasing observe that
$$V_{2j}'(\mu) = v_2(g(\frac \mu {\mu_1})) \cdot g'(\frac \mu {\mu_1}) \cdot \frac 1 {\mu_1}$$
which is positive since $h$ is non-negative and increasing and therefore $g$ is also increasing. Thus, $V_{2j}$ is increasing.

Similarly, for concavity notice that
$$ V_{2j}'' (\mu) =  \left( v_2'(g(\frac \mu {\mu_1})) \cdot g'(\frac \mu {\mu_1})^2 + 
                      v_2(g(\frac \mu {\mu_1})) \cdot g''(\frac \mu {\mu_1}) \right) \frac 1 {\mu_1^2}.$$
The first term is negative since $v_2$ is a decreasing function, and the second term is also negative since $g$ is the inverse of a positive convex function. 

The proof for $V_{1j}$ is analogous and is therefore omitted.
\end{proof}

\begin{proof}[Proof of \Cref{lem:marginal_cost_lem:fpa_spa}]
The result is a direct application of \Cref{lem0:marginals}. Consider the problem of advertiser $2$ given bids $(\mu_{1}^F,\mu_{1}^S)$. For the SPA platform the marginal cost is always the bid multiplier so 
$MC_{2S}(\mu) = \mu$. 

For the FPA platform we have that $MC_{iF}(\mu) = \mu + \frac{V_{iF}(\mu)}{V_{iF}'(\mu)}$. For advertiser 2 we have that $V_{2F}(\mu) = \int_0^{g(\mu/\mu_{1}^F)} v_2(z)dz$ where $g = h^{-1}$ so that $g(\mu, \mu_{1}^F)$ is the threshold query $q_F(\mu) := q(\mu, \mu_{i'F})$. Then, $V_{2F}(\mu) = \int_0^{q_F(\mu)} v_2(z) dz$ and 
\begin{align*}
    V_{2F}'(\mu) & =  v_2(q_F(\mu)) \cdot g'(q_F(\mu)) \cdot  \frac 1 {\mu_{1}^F} \\
    & =  v_2(q_F(\mu)) \cdot \frac 1 {h'(q_F(\mu))}  \cdot  \frac 1 {\mu_{1}^F} \\
    & = v_2(q_F(\mu)) \cdot \frac {v^2_2(q_F(\mu))}{v'_1(q_F(\mu)) v_2(q_F(\mu)) - v'_2(q_F(\mu)v_1(q_F(\mu)))} \cdot  \frac 1 {\mu_{1}^F} \\
    & = \frac{v_2(q_F(\mu))}{v_1(q_F(\mu))\mu_{1}^F} \cdot \frac {v_2(q_F(\mu))}{\eta_1(q_F(\mu)) + \eta_2(q_F(\mu))}\\
    & = \mu \cdot \frac {v_2(q_F(\mu))}{\eta_1(q_F(\mu)) + \eta_2(q_F(\mu))}.
\end{align*}
The first equality comes by taking the derivative over the integral. The second equality comes by using the derivative of the inverse function $h^{-1}$. The third equality is from the definition of $h(q)=v_1(q)/v_2(q)$. The fourth equality is a simple algebraic manipulation. The last equality comes from the definition of $q_F(\mu)$ as the threshold query when advertisers bid using $\mu_{1}^F$ and $\mu$. Hence, we conclude that  $MC_{2F}(\mu) = \mu \cdot \ela_2(q_F(\mu))$.   

The proof for advertiser 1 is identical and is therefore omitted.
\end{proof}

\begin{proof}[Proof of \Cref{theo:spa_fpa__subgame}]
From \Cref{lem:marginal_cost_lem:fpa_spa} we have that the system of equations~\eqref{eq:intrabidder1}-\eqref{eq:tcpa-final-spa-fpa} corresponds to the system of equations~\eqref{eq:marg_equalization} given bids $(\mu^*_{2F},\mu^*_{2S})$ for advertiser $1$. Since $(\mu^*_{1F},\mu^*_{1S})$  solves the system with a non-zero solution and from \Cref{lem:subgame_fpa_spa} the endogenous bid landscape satisfies Assumption~\ref{ass:vjs}. Therefore, from \Cref{theo:marg_sufficient_condition} we have that $(\mu^*_{1F},\mu^*_{1S})$ are the optimal bids for advertiser $1$ given bids $(\mu^*_{2F},\mu^*_{2S})$. The same reasoning applies to advertiser $2$ which implies that $(\mu^*_{1F},\mu^*_{1S},\mu^*_{2F},\mu^*_{2S})$ is an equilibrium for the subgame.

For uniqueness, consider an equilibrium $(\mu_{1}^F,\mu_{1}^S,\mu_{2}^F,\mu_{2}^S)$ of the subgame. We assert that bidders cannot be bidding zero on a platform. If bidding only on the FPA platform, then bidding a bid multiplier of $1$ on the SPA platform dominates bidding $0$. Under our equilibrium refinement, this cannot happen in equilibrium. If bidding only on the SPA platform, bidding a bid multiplier of $1$ on the FPA platform is feasible and dominates bidding $0$. This again contradicts our equilibrium refinement. Thus, bid multipliers are positive. 
Second, because $v_1(0)=0$ we have that for $(\mu_{2}^F,\mu_{2}^S)$ the bid landscape of advertiser is such that $\hm_{1S}=\hm_{1F} = \infty$. Analogously, for the landscape of advertiser $2$ we have that $\hm_{2S}=\hm_{2F} = \infty$. From \Cref{coro:vj_conditions} we conclude that in any equilibrium both advertisers need to simultaneously solve their system of marginal cost equations, which corresponds to the system of equations~\eqref{eq:intrabidder1}-\eqref{eq:tcpa-final-spa-fpa}. The uniqueness of solution to this system of equations implies that the equilibrium is unique.
\end{proof}

\section{Missing Proofs from Section~\ref{sec:necessary_condition}}
\subsection{Uniform Bid Multipliers Across Platforms} \label{sec:unif_mult_across_plat}




\begin{claim}
Any strategy where advertiser $i$ uses a multiplier $\mu_i < 1$ is weakly dominated by a strategy where they use a multiplier of $\mu_i = 1$.  
\end{claim}

\begin{proof}
If an advertiser bids using $\mu_i<1$, the total payment is at most $\sum_{j}\int_{S_{ij}}\mu_i v_i(q)$ which is strictly less than their value $\sum_{j}\int_{S_{ij}}v_i(q)$. By using $\mu_i=1$ the advertiser could purchase weakly more queries than with $\mu_i<1$ without violating their ROI constraint.
\end{proof}

\begin{theorem}
If advertisers are using undominated strategies, then
the unique equilibrium of the platform-advertiser game under uniform bidding is for the platforms to announce FPA to sell their queries and for advertisers to submit multipliers $\mu_i = 1$. Furthermore, the total revenue across all platforms satisfy that $\sum_j \Rev_j^* = W^*$.
\end{theorem}

\begin{proof}
Let $W^*_j = \int_{S_j}\sum_{i\in I}\max_{i\in I}v_{i,j}(q)dq $, which is the fraction of liquid welfare that platform $j$ contributes to the liquid welfare $W^*$. Recall that $W^* = \sum_{j\in J} W^*_j$.

We claim that in any equilibrium, the revenue platform $j$ gets, $\Rev_j$, satisfies that $\Rev_j\geq W^*_j$.
Indeed because advertisers are using undominated bidding stategies, if platform $j$ uses an FPA then the revenue it would get for each query $q\in S_j$, which is $\max_{i\in I}\mu_i v_{ij}(q)$, is no less than $\max_{i\in I}v_{ij}(q)$ since $\mu_i\geq 1$ for all $i\in I$. Thus, $\Rev_j\geq W^*_j$. Combining this with Remark~\ref{rem:LW} we get that $\Rev_j= W^*_j$.

To conclude observe that if valuations are in generalized position (i.e., valuations functions are not fully correlated within advertisers), then there is no $\mu_i\geq 1$ such that
$$\int_{S_j}\mbox{2nd-max}_{i\in I}\{\mu_i v_{ij}(q)\}dq =\int_{S_j}\max_{i\in I} {v_{ij}(q)}dq .$$
Hence, FPA is the unique equilibrium of the game.
\end{proof}

\subsection{Single-Bidder, Multi-Platform Problem}\label{sec:single-bidder}

In this section we study the case of a single bidder and multiple platforms each with possibly different markets made up of static bidders, with non-uniform bids allowed. By static bidders we mean that they are single-query bidders who will bid truthfully and will not respond to changes in the auction format. The main takeaway from this section is that the unique Nash equilibrium of the game is for the platforms to announce FPA to sell their queries, and for the single strategic bidder to submit a multiplier of $\mu_i = 1$. 

\begin{theorem}
In the single-bidder, non-uniform bidding multi-platform game, the unique Nash equilibrium is for the platforms to announce FPA to sell their queries and for the bidder to use a multiplier of 1 on each channel. 
\end{theorem}

We first prove the following Claim. 

\begin{claim}
For any platform $j \in J$, announcing FPA to sell their queries weakly dominates SPA. 
\end{claim}

\begin{proof}
Let $h(x)$ be the impressions the bidder gets if they bid $x$, and $p(q) = h^{-1}(q)$ be its inverse, the price the bidder must bid in order to win all impressions up to $x$ (assume $h$ is weakly monotone increasing).

Suppose a platform has announced SPA to sell their queries.  Let $\mu_S$ be the multiplier the single strategic bidder uses in order to bid on this platform. The strategic bidder wins the first $h(\mu_S)$ queries and pays a total of  $\int_0^{h(\mu_S)} p(z) dz$. The static bidders will win the queries $q > h(\mu_S)$ and pay $\mu_S$ for each. 

Suppose the platform announced FPA to sell their queries, and let $\mu_F$ be the multiplier the single strategic bidder uses in order to bid on this platform. 

By switching from SPA to FPA, the price weakly increases on all queries. For queries $q$ between $0 < q < \min\{h(\mu_F),h(\mu_S)\}$ and queries $q$ between $\max\{(h(\mu_F),h(\mu_S)\} < q$, the winner of the query is the same between both auction formats but the price increases. For the remaining queries, the winner of the query changes. If $\mu_F < \mu_S$, the price weakly increases. If $\mu_S < \mu_F$ the price remains the same. In all cases, the revenue of the platform is weakly greater announcing FPA to sell their queries compared to SPA. 
\end{proof}

We can now easily prove the main result of this section. 

\begin{proof}[Proof of \Cref{thm:single-strategic}]
Since FPA weakly dominates SPA for all platforms, the platforms will all announce FPA to sell their queries, regardless of what other platforms are doing. This in turn makes the bidder's problem simple. Submitting a multiplier $\mu_j \neq 1$ for platform $j$ is weakly dominated by submitting $\mu'_j = 1$. Submitting a multiplier $\mu_j > 1$ will result in overpaying for all the queries the strategic bidder wins. Submitting a multiplier of $\mu_j < 1$ is weakly dominated by submitting a multiplier of $\mu'_j = 1$ since this wins weakly more queries while satisfying the \tcpa constraint. 
\end{proof}

\section{Missing Proofs from Section~\ref{sec:inefficiency_free} }\label{sec:proof_inefficiency_free}

We begin with the setting of two platforms with asymmetric market share. We denote $\gamma\in [0,1]$ the market share of platform $1$ (so the market share of platform $2$ is $1-\gamma$).

\subsection{Two Platforms with Asymmetric Market Share}\label{subsec:mirror_two_platform}

We will compute the advertisers' equilibrium and the platforms' revenue for each profile of the 2-player game between platforms. For ease of notation let $L_1=\int_0^{q_{\text{eff}}} v_1(q)dq, H_1=\int_{q_{\text{eff}}}^1 v_1(q)dq$, $L_2=\int_{q_{\text{eff}}}^1 v_2(q)dq$, $H_2=\int_0^{q_{\text{eff}}} v_2(q)dq$. Recall that $q_{\text{eff}}$ is the unique (as $v_1$ is strictly increasing while $v_2$ is strictly decreasing) number such that $v_1(q_{\text{eff}})=v_2(q_{\text{eff}})$. We notice that $L_1<H_1$ and $L_2<H_2$.

The following claim directly follows from the fact that for any auction format outcome, the allocation at advertisers' equilibrium is efficient.

\begin{claim}\label{claim:mirror_same_multiplier}
For any auction format outcome, the advertisers' equilibrium satisfies that $\mu_{1j}=\mu_{2j}$ for all platform $j$. 
\end{claim}
\begin{proof}
For any auction format outcome, $q_{\text{eff}}$ satisfies that $\mu_{1j}v_1(q_{\text{eff}})=\mu_{2j}v_2(q_{\text{eff}})$ for each platform $j$ since $q_{\text{eff}}$ is the threshold where the allocation switches from advertiser 1 to advertiser 2. By definition $q_{\text{eff}}$ also satisfies $v_1(q_{\text{eff}})=v_2(q_{\text{eff}})$. Thus $\mu_{1j}=\mu_{2j}$ for all $j$.
\end{proof}

Now we compute the advertisers' equilibrium for each auction outcome.

\paragraph{Case 1: (SPA, FPA)} By \Cref{claim:mirror_same_multiplier}, denote $\mu^F, \mu^S$ the multipliers for both advertisers on the FPA, SPA platforms respectively. Then by \Cref{theo:spa_fpa__subgame} the following equations are satisfied:

\begin{align}
    &\text{(Target 1)} \int_{q_{\text{eff}}}^1\mu^S\gamma v_2(q)dq + \int_{q_{\text{eff}}}^1\mu^F(1-\gamma)v_1(q)dq = \int_{q_{\text{eff}}}^1\gamma v_1(q)dq + \int_{q_{\text{eff}}}^1 (1-\gamma)v_1(q)dq \label{equ:mirror_eq1}\\
    &\text{(Target 2)}
    \int_{0}^{q_{\text{eff}}}\mu^S\gamma v_1(q)dq + \int_0^{q_{\text{eff}}}\mu^F(1-\gamma)v_2(q)dq = \int_0^{q_{\text{eff}}}\gamma v_2(q)dq + \int_{0}^{q_{\text{eff}}} (1-\gamma)v_2(q)dq \label{equ:mirror_eq2}\\
    &\text{(Equalizing marginals 1) } \frac{\mu^S}{\mu^F} = \ela_1(q_{\text{eff}})= 1+\frac{\eta_1(q_{\text{eff}})+\eta_2(q_{\text{eff}})}{\gamma v_1(q_{\text{eff}})}\int_{q_{\text{eff}}}^1\gamma v_1(q)dq \label{equ:mirror_eq3}\\
    &\text{(Equalizing marginals 2) } \frac{\mu^S}{\mu^F} = \ela_2(q_{\text{eff}})= 1+\frac{\eta_1(q_{\text{eff}})+\eta_2(q_{\text{eff}})}{(1-\gamma) v_2(q_{\text{eff}})}\int_0^{q_{\text{eff}}}(1-\gamma) v_2(q)dq \label{equ:mirror_eq4}
\end{align}


\begin{claim}\label{claim:mirror_ela_equal}
$\ela_1(q_{\text{eff}})=\ela_2(q_{\text{eff}})$. Moreover, $\int_{q_{\text{eff}}}^1 v_1(q)dq = \int_{0}^{q_{\text{eff}}}v_2(q)dq$, $\int_0^{q_{\text{eff}}} v_1(q)dq = \int_{q_{\text{eff}}}^1 v_2(q)dq$, i.e. $H_1=H_2$ and $L_1=L_2$.
\end{claim}
\begin{proof}
$\ela_1(q_{\text{eff}})=\ela_2(q_{\text{eff}})$ and $H_1=H_2$ directly follows from \Cref{equ:mirror_eq3} and \Cref{equ:mirror_eq4}, since $v_1(q_{\text{eff}})=v_2(q_{\text{eff}})$. For the remaining equality $L_1=L_2$, we notice that
\begin{align*}
    \text{\Cref{equ:mirror_eq1}} \Leftrightarrow \mu^S\gamma\cdot L_2+\mu^F(1-\gamma)\cdot H_1 = H_1\\
    \text{\Cref{equ:mirror_eq2}} \Leftrightarrow \mu^S\gamma\cdot L_1+\mu^F(1-\gamma)\cdot H_2 = H_2\\
\end{align*}

Subtracting the second equality from the first equality implies $L_1=L_2$ since $H_1=H_2$.
\end{proof}

By \Cref{claim:mirror_ela_equal}, we denote $\ela=\ela_1(q_{\text{eff}})=\ela_2(q_{\text{eff}})$, $H=H_1=H_2$ and $L=L_1=L_2$.

\begin{lemma}\label{lem:sym:spfp}
There exists a unique advertisers' equilibrium with multipliers $(\mu^S, \mu^F)$ in the (SPA, FPA) profile such that
\begin{itemize}
    \item $\frac{\mu^S}{\mu^F}=\ela,$
    \item $\mu^S \gamma L+\mu^F (1-\gamma) H = H.$
\end{itemize}
\end{lemma}
\begin{proof}
Both conditions in the statement directly follow from Equations (1) - (4) and \Cref{claim:mirror_ela_equal}.
It remains to verify that there exists $\mu^S, \mu^F > 0$ that satisfy both conditions. In fact, we have
$$1/\mu^F=\gamma\cdot \ela \cdot \frac{L}{H} + (1-\gamma) > 0$$
and $\mu^S = \ela\cdot \mu^F > 0$.
The lemma then follows from \Cref{theo:spa_fpa__subgame}.
\end{proof}

 The revenue of platform 1 (SPA) is $\int_{0}^{q_{\text{eff}}} \gamma \mu^Sv_1(q)dq+\int_{q_{\text{eff}}}^1\gamma \mu^S v_2(q)dq=2\mu^S\gamma L$. The revenue of platform 2 (FPA) is $\int_{0}^{q_{\text{eff}}} (1-\gamma) \mu^Fv_2(q)dq+\int_{q_{\text{eff}}}^1(1-\gamma) \mu^F v_1(q)dq=2\mu^F(1-\gamma) H$.  By \Cref{lem:sym:spfp} the total revenue of both platforms is $2H$.

\paragraph{Case 2: (FPA, SPA)} This case is analogous to the previous case by swapping the market share parameter. We have the following lemma analogous to \Cref{lem:sym:spfp}. The proof is omitted.
\begin{lemma}\label{lem:sym:fpsp}
There exists a unique advertisers' equilibrium with multipliers $(\tilde\mu^S, \tilde\mu^F)$ in the (FPA, SPA) profile such that
\begin{itemize}
    \item $\frac{\tilde\mu^S}{\tilde\mu^F}=\ela,$
    \item $\tilde\mu^S (1-\gamma) L+\tilde\mu^F \gamma H = H.$
\end{itemize}
\end{lemma}

\paragraph{Case 2: (FPA, FPA)} We know that advertisers will set multipliers $\mu_{ij} = 1$. In both platforms, advertiser 1 wins the query when $q\in [q_{\text{eff}}, 1]$. Now the revenue of platform 1 is $\int_0^{q_{\text{eff}}}\gamma v_2(q)dq+\int_{q_{\text{eff}}}^1\gamma v_1(q)dq=2\gamma \cdot H$. Similarly the revenue of platform 2 is $2(1-\gamma)\cdot H$.

\paragraph{Case 2: (SPA, SPA)} Now both advertisers will set the multipliers such that the payment meets their target, i.e. $\gamma\int_{q_{\text{eff}}}^1 \mu v_2(q)dq+(1-\gamma)\int_{q_{\text{eff}}}^1 \mu v_2(q)dq = \gamma\int_{q_{\text{eff}}}^1 \mu v_1(q)dq+(1-\gamma)\int_{q_{\text{eff}}}^1 \mu v_1(q)dq \Longleftrightarrow \mu = H/L$.
The revenue is $\gamma\int_{q_{\text{eff}}}^1 \mu v_2(q)dq+\gamma\int_0^{q_{\text{eff}}} \mu v_1(q)dq = 2\gamma \cdot H$ for platform 1. Similarly the revenue of platform 2 is $2(1-\gamma)\cdot H$.

Finally, we can write the payoffs for each platform under all possible action profiles. In particular, note that the welfare of every action profile is $2H$. So long as the payoff matrix is not degenerate, the Nash equilibria of this game will be exactly one of (SPA, SPA) or (FPA, FPA). Surprisingly, which of these two profiles is the Nash equilibrium will depend only on the shape of the valuation and not on the market share $\gamma$. 

  \begin{table}[h!]
  \centering
    \setlength{\extrarowheight}{8pt}
    \begin{tabular}{cc|c|c|}
      & \multicolumn{1}{c}{} & \multicolumn{2}{c}{Ch 2}\\
      & \multicolumn{1}{c}{} & \multicolumn{1}{c}{SPA}  & \multicolumn{1}{c}{FPA} \\\cline{3-4}
      \multirow{2}*{Ch 1}  & SPA & $(2\gamma H, 2(1-\gamma) H)$ & $(2\mu^S\gamma L, 2\mu^F(1-\gamma)H)$ \\\cline{3-4}
      & FPA & $(2\tilde{\mu}^F\gamma H, 2\tilde{\mu}^S(1-\gamma) L)$ & $(2\gamma H, 2(1-\gamma) H)$ \\\cline{3-4}
    \end{tabular}
  \end{table}

\begin{theorem}\label{lem:symeq}
Let $Q=\ela\cdot\frac{L}{H}$ be the parameter that only depends on $v_1$ and $v_2$ (independent of the market share $\gamma$). If $Q>1$, then (SPA, SPA) is the only NE of the game. If $Q<1$, then (FPA, FPA) is the only NE. If $Q=1$, then all four profiles of the game have the same outcome.
\end{theorem}

Before going to the proof, we notice that
$$\compet=1-\frac{\int_0^1 |v_1(q)-v_2(q)|dq}{\int_0^1 \max\{v_1(q), v_2(q)\}dq}=\frac{\int_0^{q_{\text{eff}}}v_1(q)dq+\int_{q_{\text{eff}}}^1 v_2(q)dq}{\int_0^{q_{\text{eff}}}v_2(q)dq+\int_{q_{\text{eff}}}^1 v_1(q)dq}=\frac{L}{H}$$

So \Cref{lem:symeq} is a restatement of \Cref{thm:symeq} and \Cref{thm:market_share_sym} when there are 2 platforms.

\begin{proof}[Proof of \Cref{lem:symeq}]

Let $a=2\gamma H, b=2\tilde{\mu}^F\gamma H, c=2\mu^S\gamma L$, the game can be rewritten as the following table:

  \begin{table}[H]
  \centering
    \setlength{\extrarowheight}{8pt}
    \begin{tabular}{cc|c|c|}
      & \multicolumn{1}{c}{} & \multicolumn{2}{c}{Ch 2}\\
      & \multicolumn{1}{c}{} & \multicolumn{1}{c}{SPA}  & \multicolumn{1}{c}{FPA} \\\cline{3-4}
      \multirow{2}*{Ch 1}  & SPA & $(a, 2H-a)$ & $(c, 2H-c)$ \\\cline{3-4}
      & FPA & $(b, 2H-b)$ & $(a, 2H-a)$ \\\cline{3-4}
    \end{tabular}
  \end{table}

{Observe that in order for (SPA, SPA) to be a Nash equilibrium we need that $a > b$, $2H-a > 2H-c$ (or $c > a$). Analogously, in order for (FPA, FPA) to be a Nash equilibrium we need that $c < a$ and $2H- a > 2H - b$ (or $a < b$). We will now show that $Q < 1 \iff b > a \iff a > c$.}

\begin{claim}
{$Q < 1 \iff b > a \iff a > c$.}
\end{claim}

\begin{proof}
By \Cref{lem:sym:fpsp},
\begin{align*}
(1-\gamma) H Q \tilde{\mu}^F +\tilde{\mu}^F\gamma H & = H \\
1/\tilde{\mu}^F & =Q (1-\gamma)+\gamma.
\end{align*}
Similarly by \Cref{lem:sym:spfp}, 

\begin{equation}\label{equ:mirror_mu_F}
1/\mu^F=Q \gamma+(1-\gamma)
\end{equation}

By definition,
\begin{align*}
a<b & \iff \tilde{\mu}^F>1 \\ 
& \iff Q(1-\gamma)+\gamma<1 \\ 
& \iff Q < 1
\end{align*}
Moreover,

\begin{align*}
a>c & \iff \mu^S < H/L \\ 
& \iff 1/\mu^F > \frac{L}{H}\cdot \ela = Q \\
& \iff Q\gamma+(1-\gamma)>Q \\
& \iff Q < 1.
\end{align*}
Thus if $Q < 1$ we have that $b>a>c$ and if $Q > 1$ we have that $c>a>b$. 
\end{proof}

Therefore, if $Q > 1$, the only equilibrium is (SPA, SPA). If $Q < 1$, the only equilibrium is (FPA, FPA). If $Q = 1$, then the game is degenerate and all action profiles have the same payoffs. 
\end{proof}

\subsection{Equilibrium for Standard Valuations in the Mirrored Setting}\label{subsec:mirror_case_study}

We apply the result of \Cref{lem:symeq} to show that (SPA, SPA) is the only Nash equilibrium for many standard valuations. Throughout this section we focus on the mirrored setting, i.e. $v_1(q)=v_2(1-q)=v(q)$ for all $q\in [0,1]$ for some $v$. In this setting we have $q_{\text{eff}}=\frac{1}{2}$ and
$$\ela = 1+\frac{2v'(\frac{1}{2})}{v^2(\frac{1}{2})}\cdot H,\qquad Q=\ela\cdot \frac{L}{H} = \frac{L}{H}+\frac{2v'(\frac{1}{2})}{v^2(\frac{1}{2})}\cdot L$$

\subsubsection{Polynomial valuations $v(q)= \sum_{i \geq 1} \alpha_i q^{i} (\alpha_i\geq 0)$}



\begin{corollary}\label{cor:polynomial}
For polynomial valuations with positive coefficients and no constant term such that $\frac{v(q)}{v(1-q)}$ is convex, (SPA, SPA) is always the only Nash equilibrium. 
\end{corollary}

\begin{proof}
Let $v(q) = \sum_{i \geq 1} \alpha_i q^i$.

\begin{align*}
    v'(1/2) &= \sum_i \alpha_i \cdot i \cdot 2^{1-i} \\ 
    v^2(1/2) &= \left(\sum_i \alpha_i 2^{-i}\right)^2 \\  
    L &= \int_{0}^{1/2} \sum_i \alpha_i q^i dq = \sum_i \frac{\alpha_i 2^{-(i+1)}}{(i+1)}.
\end{align*}

First note that if $L = 0$, then all $\alpha_i$ must be $0$, in which case $H$ is also $0$ and the game is degenerate, hence any action profile would be a Nash equilibrium. Assume $L > 0$ for the rest of this proof. We will prove that $Q > 1$ directly by showing that $2 \cdot \frac{2v'(1/2)}{v^2(1/2)} \cdot L \geq 1$. By Cauchy-Schwarz we know that 

\begin{align*}
    v'(1/2) \cdot L \geq \left(\sum_i \frac{\alpha_i 2^{-i} \sqrt{i}}{\sqrt{i+1}} \right)^2. \\ 
\end{align*}

Multiplying both sides by 2 and observing that $2\frac{i}{i+1} \geq 1$ (for all $i \geq 1$) gives the inequality. Since $L > 0$, then $Q>1$. Finally, note that we need convexity of $v(q)/v(1-q)$ for our results to hold. 
\end{proof}


We can specialize the results to the case of monomial valuations $v(q) = q^{\alpha}$ for some $\alpha > 0$. \Cref{cor:mono} directly follows from \Cref{cor:polynomial} with the fact that $(\frac{q}{1-q})^\alpha$ is convex when $\alpha\geq 1$.

\begin{corollary}\label{cor:mono}
For monomial valuations, (SPA, SPA) is always the only Nash equilibrium. 
\end{corollary}

\subsubsection{Fixed crossing point}

Given \Cref{lem:symeq}, we can think of how the equilibrium changes for families of valuations with, say, fixed crossing points. For this section, assume $v(1/2) = 1/2$. Moreover, we assume $v(q) \geq 0, \forall q$. 

\begin{corollary}
For linear valuations with $v(1/2) = 1/2$ and $v(q)\geq 0$, (SPA, SPA) is the unique Nash equilibrium. 
\end{corollary}

\begin{proof}
We will again show directly that $Q \geq 1$. We consider two sub-cases. Let $v(q) = \alpha q + \beta$ be the valuation, and since $v(1/2)=1/2$ we know that $\beta = 1/2 - \alpha/2$. Also, note that $L + H = 1/2$. 

Since $v(q) \geq 0$, $0 \geq \alpha \geq 1$. Therefore we can compute $L$ directly. 

\begin{align*}
    L &= \int_{0}^{1/2} (\alpha q + 1/2 - \alpha/2) dq \\ 
    & = \alpha/8 + 1/4- \alpha/4 =  1/4-\alpha/8.\\  
\end{align*}

Thus, $Q = \frac{1/4 - \alpha/8}{1/4+\alpha/8}+2\frac{\alpha}{1/4}(1/4-\alpha/8) = \frac{2-\alpha}{2+\alpha} + \alpha \cdot (2-\alpha)$. It is easy to verify that for $0 \leq \alpha \leq 1$, $Q \geq 1$ with equality only when $\alpha = 0$.
\end{proof}
    
\subsubsection{Fixed slope} 

Similar to the previous case, we can consider what happens with linear functions such that $v'(1/2) = \alpha$ for some fixed $\alpha$. Without loss of generality, assume $\alpha = 1$. Let $v(q) = q + \beta$. We further assume that $v(q) \geq 0 \forall q$, thus $\beta \geq 0$.  

\begin{corollary}
For linear functions with $v'(1/2) = 1$, the unique equilibrium will be (SPA, SPA). 
\end{corollary}

\begin{proof}
Then $L = \frac{2\beta + 1/2}{4}$, $H = \frac{2\beta+3/2}{4}$. Therefore, $Q = \frac{2\beta+1/2}{2\beta+3/2}+2\frac{2\beta+1/2}{(\beta+1/2)^2}$. It is easy to verify that for $\beta \geq 0$, $Q > 1$.


\end{proof}

\subsubsection{Exponential valuations}

\begin{corollary}
For exponential valuations $v(q) = e^{\alpha q}-1$ with $\alpha > 0$, it is always a NE to have (SPA, SPA). 
\end{corollary}

\begin{proof}
We will prove that the second term of $Q$ is already greater than $1$ for all $\alpha > 0$.

\begin{align*}
    L & = \int_0^{1/2} (e^{\alpha x} - 1) dx \\
    & = \frac{e^{\alpha/2}}{\alpha}-1/2. 
\end{align*}

Furthermore, $v'(1/2) = \alpha (e^{\alpha/2})$, $v^2(1/2) = (e^{\alpha/2}-1)^2$. Then the condition becomes 

\begin{align*}
    \frac{e^{\alpha/2}(2e^{\alpha/2}-\alpha)}{(e^{\alpha/2}-1)^2}.
\end{align*}

It is easy to verify that this expression is at greater than 1 for $\alpha > 0$. 

\end{proof}

\subsection{Multiple Platforms} \label{subsec:mirror_multi_platforms}

In this section we prove \Cref{thm:market_share_sym} using the results we developed in \Cref{subsec:mirror_two_platform}.

\begin{proof}[Proof of \Cref{thm:market_share_sym}]
Fix any platform $j$ and the auction choices of the other platforms. We are going to analyze platform $j$'s best response at the current outcome. Let $\gamma'$ be the total market share of all other platforms that choose SPA. Then the total market share of all other platforms that choose FPA is $1-\gamma'-\gamma_j$. We compute as follows the advertiser equilibrium and revenue of platform $j$ if the platform chooses SPA and FPA respectively.

To achieve this, consider any auction outcome and denote $\gamma\in [0,1]$ the total market share of the platforms that choose SPA. By \Cref{claim:mirror_same_multiplier}, advertisers will use the same multiplier on all platforms that chooses SPA, which we denote as $\mu^S(\gamma)$. Similarly denote $\mu^F(\gamma)$ the multipliers on the FPA platforms. We notice that by merging all SPA platforms and FPA platforms respectively, the advertiser equilibrium is equivalent to the one of the (SPA, FPA) profile in the two-platform setting. By \Cref{lem:sym:spfp}, $(\mu^S(\gamma), \mu^F(\gamma))$ satisfy that 
\begin{itemize}
    \item $\frac{\mu^S(\gamma)}{\mu^F(\gamma)}=\ela,$
    \item $\mu^S(\gamma)\cdot \gamma L+\mu^F(\gamma)\cdot (1-\gamma) H = H.$
\end{itemize}
Or equivalently (by \Cref{equ:mirror_mu_F}), 
$$\mu^F(\gamma) = \frac{1}{Q\gamma+1-\gamma},\qquad \mu^S(\gamma)=\ela \cdot \mu^F(\gamma)=\mu^F(\gamma)\cdot Q\cdot \frac{H}{L}$$

\paragraph{The platform chooses SPA} We notice that SPA has a total market share of $\gamma'+\gamma_j$. Thus both advertisers use multiplier $\mu^S(\gamma'+\gamma_j)$ for platform $j$. Recall that in the advertiser equilibrium, advertiser $1$ wins query $q\in[q_{\text{eff}}, 1]$ and advertiser $2$ wins query $q\in[0, q_{\text{eff}})$. The revenue of platform $j$ is 
$$\mathcal{R}_S=\gamma_j\cdot \int_0^{q_{\text{eff}}}\mu^S(\gamma'+\gamma_j)\cdot v_1(q)dq+\gamma_j\cdot \int_{q_{\text{eff}}}^1\mu^S(\gamma'+\gamma_j)\cdot v_2(q)dq=2\gamma_j\cdot \mu^S(\gamma'+\gamma_j)\cdot L.$$

\paragraph{The platform chooses FPA} Now SPA has a total market share of $\gamma'$. Thus both advertisers use multiplier $\mu^F(\gamma')$ for platform $j$. The revenue of platform $j$ is 
$$\mathcal{R}_F=\gamma_j\cdot \int_0^{q_{\text{eff}}}\mu^F(\gamma')\cdot v_2(q)dq+\gamma_j\cdot \int_{q_{\text{eff}}}^1\mu^F(\gamma')\cdot v_1(q)dq=2\gamma_j\cdot \mu^F(\gamma')\cdot H.$$

We now compare the revenue of both choices:
\begin{align*}
\frac{\mathcal{R}_S}{\mathcal{R}_F}&=\frac{\mu^S(\gamma'+\gamma_j)}{\mu^F(\gamma')}\cdot \frac{L}{H}=Q\cdot \frac{\mu^F(\gamma'+\gamma_j)}{\mu^F(\gamma')}=Q\cdot \frac{Q\gamma'+1-\gamma'}{Q(\gamma'+\gamma_j)+1-\gamma'-\gamma_j}\\
&=Q\cdot \left(1-\frac{(Q-1)\cdot \gamma_j}{Q(\gamma'+\gamma_j)+1-\gamma'-\gamma_j}\right)
\end{align*}
Thus
$$\mathcal{R}_S>\mathcal{R}_F\Leftrightarrow \frac{(Q-1)\cdot \gamma_j}{Q(\gamma'+\gamma_j)+1-\gamma'-\gamma_j}<\frac{Q-1}{Q}\Leftrightarrow (Q-1)\cdot (1-\gamma_j+(Q-1)\gamma') > 0$$

We notice that $1-\gamma_j+(Q-1)\gamma'>1-\gamma_j-\gamma'\geq 0$ since $Q>0$ and $\gamma'+\gamma_j\leq \sum_k \gamma_k=1$. Thus $\mathcal{R}_S>\mathcal{R}_F\Leftrightarrow Q>1$. This finishes the proof: If $Q>1$, then SPA dominates FPA for any platform $j$ and any auction profile of the other platforms. Thus there is a unique platform equilibrium where every platform chooses SPA. On the other hand, if $Q<1$, then FPA is a dominate strategy for any platform. Thus the unique Nash equilibrium of the game is all platforms choosing FPA.
\end{proof}

\section{Missing Proofs from Section~\ref{sec:linearvsconstant}}
\label{app:linearvsconstant}

For this case, we can compute optimal bids for each of the strategy pairs the platforms might employ. This in turn allows us to determine the Nash equilibria of the platform-advertiser game as a function of $\alpha$. Before we can state the main result of this section, we will restrict our attention to a range of $\alpha$. As it will turn out, $\alpha$ not in this range are easy to solve and strategically not as compelling. 

\begin{proof}[Proof of Claim~\ref{cl:alpharange}]
For values of $\alpha \not \in (2,4)$, we claim that the (SP, SP) subgame will have an equilibrium where a single advertiser wins all or virtually all queries. Because of this effective lack of advertiser competition, we will rule out these values of $\alpha$. 

Regardless of $\alpha$, it will turn out that a dominant strategy for the second advertiser is to set a bid multiplier of $\mu_2 = 2$. This is because for any queries they win they will be paying $\int_0^{q_S} \mu_1 q dq$, where $q_S$ satisfies $\mu_1 q_S = 2$. Therefore, they will at most $\mu_1 q_S^2/2 \leq q_S$, which is the total value for the queries they would win. Setting a multiplier lower than $2$ is weakly dominated by setting a multiplier of $2$ since they win weakly more queries while not violating their target constraint. Setting a multiplier higher than $2$ is dominated since they will be violating their target constraint. 

First we consider the case where $\alpha \leq 2$. Since $\mu_2 = 2$, then the price on any impressions the first advertiser wins will be $2 \leq \alpha$. Therefore, winning any queries will cause the first advertiser to violate their target constraint. Because of this, the first advertiser will set $\mu_1 < 2$. Whatever $\mu_1$ is, the second advertiser will win all queries and still satisfy their target constraint. 

Next we consider the case where $\alpha \geq 4$. Since $\mu_2 = 2$, again the first advertiser faces a price of $2$ on any queries they win. Their total payment is $2 (1-q_S)$, and their total value for these queries is $\alpha/2 (1-q_S^2)$. If $\alpha \geq 4$, then the value will always exceed the cost. Therefore, they are willing to set a bid multiplier of $\mu_1 = M$ for any $M >> 2$, which would cause them to win virtually all queries. Another way to phrase this last result is that for any $\mu_1 = M >> 2$, setting a multiplier of $M' > M$ is a better strategy since it will win more queries while satisfying the target constraint. Therefore, when $\alpha \geq 4$ there is no equilibrium.
\end{proof}


The proof of Theorem~\ref{thm:mainconstantvslinear} relies on the analysis of the three different subgames -- (FPA, FPA), (SPA, SPA) and (FPA, SPA). We first compute the equilibrium bid multipliers for each subgame. These in turn define the revenue (as a function of $\alpha$) that each of the platform gets. Once we have all payoffs, we can derive conditions under which the action profiles are equilibria. As it turns out, there exist values of $\alpha$ that support all possible pure Nash equilibria in this game. 

The first subgame where both platforms use FPA is easy to analyze: advertisers will use bid multipliers of $1$, platforms will allocate efficiently and achieve a combined revenue equal to the liquid welfare.

\begin{claim}\label{cl:FPAlinear}
In the (FPA, FPA) subgame, the advertisers will bid truthfully and the allocation will be efficient. The revenue of each platform in this case is $(\alpha+\frac{1}{\alpha})/2$. 
\end{claim}

\begin{proof}
 Because platforms are identical and advertisers are being first-priced, then each advertiser uses a bid multiplier $\mu_i^F=1$ in equilibrium. To see this, fix $\mu_1^F=1$ and consider possible responses by the second advertiser. Using a bid multiplier $\mu_2^F>1$ will result in an allocation where aggregate spend is greater than aggregate value. Using a bid multiplier $\mu_2^F < 1$ is dominated since the advertiser can raise their bid to $1$ and win weakly more queries while ensuring that their aggregate spend is not greater than their aggregate value. Even though the bidders are not identical in this case, the same reasoning holds if we fix $\mu_2^F = 1$ and compute a best, undominated response for advertiser $1$. 
Thus, each platform extracts the full liquid welfare from the advertisers. Finally, observe that $q^1_F = q^2_F = \frac {1}{\alpha}.$ The revenue the each platform gets is given by  $$1 + \frac{1}{2} (\alpha-1)(1-q^1_{F}) = \frac{\alpha^2+1}{2\alpha}.$$
\end{proof}

The inefficiencies (and intricacies) of this family of valuations begin with the subgame where both platforms use SPA. In this case we first show that advertiser 2 will always bid with a multiplier of $2$. This, combined with other optimality conditions, allows us to pin down the optimal multiplier for advertiser 1, as  well as the overall revenue of the platforms as a function of $\alpha$. 

\begin{claim}\label{cl:SPAlinear}
In the (SPA, SPA) subgame, advertiser 2 will use a bid multiplier of $\mu_2 = 2$, advertiser 1 will use a bid multiplier of $4/\alpha - 1$. The revenue of each platform in this case is $3-\frac{4}{\alpha}$. 
\end{claim}

\begin{proof}
Since the platforms are indistinguishable, we will solve for a symmetric equilibrium where advertisers use the same multipliers for both platforms, and both platforms get the same revenue. The bid multipliers $\mu_i^S$ are chosen so that both advertisers meet their target constraint. There are three equations that govern the unknowns in this setting. 

\begin{align*}
    \mu_2^S & = \mu_1^S \alpha q_S, \\
    \mu_2^2(1-q_S) & = \int_{q_S}^{1} \alpha q dq = \alpha \frac{(1-q_S^2)}{2}, \\ 
    q_S &= \int_{0}^{q_S} \mu_1^S \alpha q dq = \mu_1^S \alpha \frac{q_S^2}{2}.\\  
\end{align*}

The first equation establishes that $q_S$ is the point at which the first advertiser becomes the highest bidder for the first time. The second equation ensures that the spend by the first advertiser (which depends on the second advertiser's multiplier) equals their value for the acquired queries. The last equation does the same for the second advertiser (whose spend depends on the first advertiser's multiplier). From the first and last equation we can infer directly that $\mu_2^S = 2$. This in turn implies that $q_S = \frac{4}{\alpha}-1$ and $\mu_1^S = \frac{2}{4-\alpha}$. The revenue each platform gets is 
$$q_s+2(1-q_s) = 3 - \frac{4}{\alpha}.$$


\end{proof}

Finally, the most intricate case is that where one platform uses SPA and the other uses FPA. Because the platforms are no longer symmetric, each advertiser may use a different multiplier on each platform. Interestingly, the multipliers for advertiser 2 are easy to obtain and understand: the advertiser will still use a multiplier of $2$ on the SPA platform and a multiplier of $1$ on the FPA platform. Utilizing the results from the previous sections, we are able to mathematically solve for the remaining variables in this scenario as a function of $\alpha$. However, those solutions are not as elegant to present. 

\begin{claim}\label{cl:FPASPAlinear}
In the (FPA, SPA) subgame, advertiser 2 will use a bid multiplier of $\mu_2^S = 2$ on the SPA platform and a multiplier of $\mu_2^F = 1$ on the FPA platform. The multipliers of advertiser 1 depend on $\alpha$ and $q_F$, the query where advertiser 1 starts winning queries on the FPA platform. The revenues of each platform can be computed as a function of $\alpha, q_F$ (where $q_F$ is uniquely determined by $\alpha$). 
\end{claim}

\begin{proof}

There are six variables to solve for $\mu_1^F, \mu_1^S, \mu_2^F, \mu_2^S, q_F, q_S$. We first write the equations which establish that $q_F, q_S$ are the points at which the second advertiser starts losing queries and the first advertiser starts winning them.

\begin{align}
    \mu_2^F & = \mu_1^F \alpha q_F \label{eq:FPSPqF} \\ 
    \mu_2^S & = \mu_1^S \alpha q_S \label{eq:FPSPqS} 
\end{align}

The remaining equations come from the analysis on the previous section. We just instantiate Eqs.\eqref{eq:intrabidder1}-\eqref{eq:tcpa-final-spa-fpa} for the valuations $v_1(q)=\alpha q$, $v_2(q)=1$. 

\begin{align}
    \frac{\mu_1^S}{\mu_1^F} & =  \frac{(1+q_F^2)}{2q_F^2} \label{eq:linearintrabidder}\\ 
    \frac{\mu_2^S}{\mu_2^F} & = 2 \label{eq:constantintrabidder} \\ 
    (\mu_1^F-1)\frac{(1-q_F^2)\alpha}{2} &= \frac{(1-q_S^2)\alpha}{2} - \mu_2^S(1-q_S) \label{eq:lineartarget} \\
    (\mu_2^F-1)q_F &= q_S - \alpha \frac{q_S^2 \mu_1^S}{2} \label{eq:constanttarget}  
\end{align}

Substituting Eqs.~\ref{eq:FPSPqS},~\ref{eq:constantintrabidder} into Eq.~\ref{eq:constanttarget}, we get that 

\begin{align*}
    (\mu_2^F-1) q_F = q_S - \mu_2^F q_S = q_S (1-\mu_2^F). 
\end{align*}

Because $q_S, q_F \geq 0$, the expression above can only hold if $\mu_2^F = 1$. Therefore, by Eq.~\ref{eq:constantintrabidder}, $\mu_2^S = 2$. We write the remaining variables in terms of $q_F$: 

\begin{align*}
    \mu_1^F &= \frac{1}{\alpha q_F} \\ 
    \mu_1^S &= \frac{(1+q_F^2)}{2\alpha q_F^3} \\ 
    q_S &= \frac{4q_F^3}{(1+q_F^2)}
\end{align*}
    
Finally, we can substitute all of these and write Eq.~\ref{eq:lineartarget} fully as a function of $q_F$ (and $\alpha$). 

\begin{align*}
    \left(\frac{(1-\alpha q_F)}{\alpha q_F}\frac{(1-q_F^2) \alpha}{2} \right) & = \left(1-\frac{(4q_F^3)^2}{(1+q_F^2)^2}\right) \frac{\alpha}{2} - 2\left(1-\frac{4q_F^3}{(1+q_F^2)}\right). 
\end{align*}

Rewriting $q_F =x$,

$$ 17 \alpha x^7 - 17 x^6 + 4x^5 - 17x^4 + (8-3 \alpha) x^3 + x^2 + (4-2 \alpha)x + 1 = 0. $$

Finally, we can write $\alpha$ as a function of $x$. 

\begin{align*}
    \alpha = \frac{17x^6-4x^5+17x^4-8x^3-x^2-4x-1}{17x^7-3x^3-2x}.
\end{align*}

With this expression relating $\alpha$ and $q_F$ we can empirically plot their relationship and observe that $q_F$ is monotone decreasing in $\alpha$. This in particular implies that as $\alpha$ increases, advertiser $1$ wins more queries in the FPA platform.

\begin{claim}\label{cl:alphamonotone}
For the (FPA, SPA) subgame, $q_F$, the point at which advertiser $1$ starts winning queries in the FPA platform, is strictly decreasing in $\alpha$. 
\end{claim}

We can now compute the revenue each platform gets. The platform running First Price gets revenue 

\begin{align*}
    \mu_2^F q_F + \int_{q_F}^{1} \mu_1^F \alpha q dq & = q_F + \frac{\alpha (1-q_F^2)}{2 \alpha q_F} \\ 
    & = \frac{1+q_F^2}{2q_F}. 
\end{align*}

The platform running Second Price gets revenue 

\begin{align*}
    \int_0^{q_S} \mu_1^S \alpha q dq + \int_{q_S}^{1} \mu_2^S dq & = \frac{1+q_F^2}{4q_F^3} \cdot  \left(\frac{4q_F^3}{(1+q_F^2)}\right)^2 + 2(1-q_S) \\
    & = \frac{4q_F^3}{(1+q_F^2)}+2\left(1-\frac{4q_F^3}{1+q_F^2}\right) = 2-\frac{4q_F^3}{1+q_F^2}. 
\end{align*}
\end{proof}

Once we have Claims~\ref{cl:FPAlinear},~\ref{cl:SPAlinear},~\ref{cl:FPASPAlinear} we can write the payoff matrix for each platform under all action profiles. With this symmetric matrix in hand, we can exactly determine the conditions (and values of $\alpha$) under which each action profile is a possible equilibrium of the game. We can now write the full payoff matrix for the ensuing game between the platforms. We are now ready to present the main result of this section, the proof of Theorem~\ref{thm:mainconstantvslinear}.


  \begin{table}[h!]
  \centering
    \setlength{\extrarowheight}{8pt}
    \begin{tabular}{cc|c|c|}
      & \multicolumn{1}{c}{} & \multicolumn{2}{c}{Platform 2}\\
      & \multicolumn{1}{c}{} & \multicolumn{1}{c}{SPA}  & \multicolumn{1}{c}{FPA} \\\cline{3-4}
      \multirow{2}*{Platform 1}  & SPA & $(3-4/\alpha, 3-4/\alpha)$ & $(2- q_S,\frac {1+q_F^2}{2q_F})$ \\\cline{3-4}
      & FPA & $(\frac {1+q_F^2}{2q_F},2- q_S)$ & $(\frac{\alpha^2+1}{2\alpha},\frac{\alpha^2+1}{2\alpha})$ \\\cline{3-4}
    \end{tabular}
        \caption{Payoff matrix for the platforms' game after solving for the advertisers' subgame equlibrium, for the valuatiosn described in Section~\ref{sec:linearvsconstant}}
    \label{table:lingame}
  \end{table}
 
\begin{proof}[Proof of Theorem~\ref{thm:mainconstantvslinear}.]

Before considering the conditions under which each profile is an equilibrium, let us review the conditions that all candidates must satisfy in order to be feasible solutions to our problem. 

\begin{align*}
    q_F & \in [0,1] \\ 
   2 &< \alpha < 4 \\
    \alpha &= \frac{17{q_F}^6-4q_F^5+17q_F^4-8q_F^3-q_F^2-4q_F-1}{17q_F^7-3q_F^3-2q_F} \in (2,4) \\ 
    & \rightarrow x_1 = .237818 \leq q_F \leq x_5 = .610502, 
\end{align*}

Let us rewrite the payoff matrix as follows. Let $x=3-4/\alpha$, $y = 2-q_S$, $z=\frac{1+q_F^2}{2q_F}$, $w=\frac{\alpha^2+1}{2\alpha}$. In these terms, (SPA, SPA) is a pure NE when $x > z$, (FPA, FPA) is a pure NE when $w > y$. Both of these conditions can happen simultaneously for a small range of values of $\alpha$, and in that case their mixture is also a Nash equilibrium of the game. If neither of these conditions holds, then (FPA, SPA), (SPA, FPA) and a mixture between these two are the only Nash equilibrium.

\begin{itemize} 
\item If $x > z \rightarrow x_2 \leq q_F \leq x_4$, where $x_2 = .287407, x_4 = .565389$. 
\item If $w > y \rightarrow x_1 \leq q_F \leq x_3$, where $x_1 = .237818, x_3 = .295027$.
\item If both $x > z \land w > y \rightarrow x_2 \leq  q_F \leq x_3$.   
\item If $z > x \land y > w \rightarrow q_F \geq x_4 \lor q_F \leq x_1$. 
\end{itemize}

Therefore, if 
\begin{itemize} 
\item $x_1 \leq q_F \leq x_2$, (FPA, FPA) is the unique Nash equilibrium, 
\item $x_2 \leq q_F \leq x_3$, then (SPA, SPA), (FPA, FPA) and a mixture between the two are all Nash equilibria, 
\item $x_3 \leq q_F \leq x_4$, (SPA, SPA) is the unique Nash equilibrium,
\item $x_4 \leq q_F \leq x_5$, then (FPA, SPA), (SPA, FPA) and a mixture of the two are all equilibria. 
\end{itemize}

Finally, because for $\alpha \in (2, 4)$, $q_F$ is monotone decreasing in $\alpha$, we get that there exist $\alpha_0 = 2 < \alpha_1 < \alpha_2 < \alpha_3 < \alpha_4 =4$ where the equilibria change as indicated above.

Therefore, if 
\begin{itemize} 
\item $4 > \alpha \geq \alpha_3 = 3.5822$, (FPA, FPA) is the unique Nash equilibrium, 
\item $\alpha_3 \geq \alpha \geq \alpha_2 = 3.52753$, then (SPA, SPA), (FPA, FPA) and a mixture between the two are all Nash equilibria, 
\item $\alpha_2 = 3.52753 \geq \alpha \geq \alpha_1 = 2.1822$, (SPA, SPA) is the unique Nash equilibrium,
\item $\alpha_1 = 2.1822 \geq \alpha > \alpha_0 = 2$, then (FPA, SPA), (SPA, FPA) and a mixture of the two are all equilibria. 
\end{itemize}

\end{proof}

\section{Missing Proofs From Section ~\ref{sec:expvsexp}}
\label{app:expvsexp}



Just like the proof of Theorem~\ref{thm:mainconstantvslinear}, the proof of Theorem~\ref{thm:expequilibria} will begin by finding the subgame equilibria for each of the three possible subgames. We will also compute each platforms' payoffs under the advertisers' equilibrium multipliers. Finally, we will compute the payoff matrix and prove numerically that SP is a dominant strategy of the game. 

\begin{claim}
\label{cl:expfpfp}
In the (FP, FP) subgame, both advertisers will use multipliers of $1$. The revenue of each platform will be $\frac{\alpha^2+1}{2}$.
\end{claim}

\begin{proof}
In this case, since both platforms are symmetric and both running first price auctions, the advertisers will set their multipliers to 1. The query $q_{\text{eff}}$ at which advertiser 2 stops winning and advertiser 1 starts winning is thus the one where $\alpha e^{-q_{\text{eff}}} = e^{-2q_{\text{eff}}}$, or $\alpha = e^{-q_{\text{eff}}}$. The revenue is thus $\int_{0}^{q_{\text{eff}}} e^{-2q} dq + \int_{q_{\text{eff}}}^{\infty} \alpha e^{-q} dq = (-e^{-2q_{\text{eff}}}+1)/2 + \alpha e^{-q*} = \alpha^2 + 1/2 - \alpha^2/2 = \frac{1+\alpha^2}{2}$, where we used the fact that $\alpha = e^{-q_{\text{eff}}}$.
\end{proof}

\begin{claim}
\label{cl:expspsp}
For $\alpha \in (1/4, 1/2)$, advertiser 1 will use a multiplier of $\mu_1^S = 2$ and advertiser 2 will use a multiplier of $\mu_2^S = \frac{2\alpha}{4\alpha -1}$. The revenue of each platform will be $(3-4\alpha)\alpha$.  
\end{claim}

\begin{proof}
In this case, since the platforms are symmetric, the advertisers will use the same multipliers $\mu_1^S, \mu_2^S$ across both platforms. The efficient $q_{\text{eff}}$ is such that 
$$ \mu_1^S \alpha e^{-q_{\text{eff}}} = \mu_2^S e^{-2q_{\text{eff}}}. $$

In other words $q_{\text{eff}}$ is such that $\frac{\mu_1^S \alpha}{\mu_2^S} = e^{-q_{\text{eff}}}$. The bidders will use multipliers so that their targets are met. We get the following two equations, one per advertiser:

$$ \int_{q_{\text{eff}}}^{\infty} \alpha e^{-q} dq = \int_{q_{\text{eff}}}^{\infty} \mu_2^S e^{-2q} dq,$$   

$$ \int_0^{q_{\text{eff}}} e^{-2q} dq = \int_0^{q_{\text{eff}}} \mu_1^S \alpha e^{-q} dq.$$   

From the first equation, $\alpha e^{-q_{\text{eff}}} = \mu_2^S e^{-2q_{\text{eff}}}/2$. Therefore, $\mu_2^S = 2 e^{q_{\text{eff}}} \alpha$. This combined with the first equation relating $\mu_1^S, \mu_2^S$ implies $\mu_1^S = 2$. From the second equation, let $x = e^{-q_{\text{eff}}}$. Rewriting the second equation gives a quadratic equation on $x$. 

$$ (1-x^2)/2 = 2 \alpha (1-x). $$

We manipulate further to get the following. 

$$ 1 - x^2 = 4 \alpha - 4 \alpha x $$
$$ 1-4\alpha = x^2 - 4 \alpha x $$ 
$$ 1-4\alpha + 4 \alpha^2 = x^2 - 4 \alpha x + 4 \alpha^2 $$ 
$$ (1-2\alpha)^2 = (x-2\alpha)^2 $$ 

This only has two sensible solutions. Either the left hand side and the right hand side have the same sign, or they have opposite signs. If they have the same sign, we get that $x = 1$ and $\mu_2^S = 2 \alpha$. This solution only makes sense if $\mu_2^S \geq 1$, i.e. $\alpha \geq 1/2$. Also $x = 1$ implies $q_{\text{eff}} = 0$. We will rule out this case since it will lead to a non-interior allocation, i.e. one advertiser will win all impressions by bidding over the other advertiser. If they don't have the same sign, we get that $x = 4\alpha - 1$. Recall that $x = e^{-q_{\text{eff}}}$. Since $q_{\text{eff}} \geq 0$, then $0 \leq x \leq 1$. Thus in the other case, we get that $0 \leq 4 \alpha - 1 \leq 1$, or $1/4 \leq \alpha \leq 1/2$ and we get that $\mu_2^S = \frac{2\alpha}{4\alpha-1}$. This proves Claim~\ref{cl:expalpha}.  

Then the revenue can be written as 

$$ \int_0^{q_{\text{eff}}} \mu_1^S \alpha e^{-q} dq + \int_{q_{\text{eff}}}^{\infty} \mu_2^S e^{-2q} dq $$ 
$$ 2 \alpha (1-e^{-q_{\text{eff}}}) + \mu_2^S (e^{-2q_{\text{eff}}})/2$$


Because $x=4\alpha-1$, this is the same as 
$$ 2\alpha (1-x) + \left(\frac{2\alpha}{x} x^2\right)/2 = 2 \alpha - \alpha x = \alpha(2-x) = \alpha(3-4\alpha).$$ 
\end{proof}

For the remaining subgame we are able to find the equations that uniquely determine the equilibrium multipliers, but we are not able to write them in a simple closed form.  


\begin{claim}
\label{cl:expfpsp}
In the (FP, SP) subgame, we present the system of equations that pins down the value of all multipliers and express the overall revenue of each platform as a function of these multipliers and $\alpha$. 
\end{claim}

\begin{proof}
In this case we need four multipliers: $\mu_1^F, \mu_1^S, \mu_2^F, \mu_2^S$, one for each advertiser, platform pair. The index denotes the advertiser, the exponent denotes whether it is the first price (F) or second price (S) platform. The elasticity of the valuations are $\nu_1(q) = |\frac{-\alpha e^{-q}}{\alpha e^{-q}}|=1$, $\nu_2(q) = |\frac{-2 e^{-2q}}{e^{-2q}}| = 2$.

From Eqs.\eqref{eq:intrabidder1}-\eqref{eq:tcpa-final-spa-fpa} we get that 

\begin{align}
\frac{\mu_1^S}{\mu_1^F} &= 1+\frac{1+2}{\alpha e^{-q^F}} \int_{q^F}^{\infty} \alpha e^{-q} dq = 4. \label{eq:expintra1} \\ 
\frac{\mu_2^S}{\mu_2^F} &= 1 + \frac{1+2}{e^{-2q^F}} \int_{0}^{q^F} e^{-2q} dq = 1 + \frac{3}{2} \frac{(1-e^{-2q^F})}{e^{-2q^F}}. \label{eq:expintra2} 
\end{align}

From Eqs.~\ref{eq:tcpa-final-spa-fpa-2},~\ref{eq:tcpa-final-spa-fpa} we get that 

\begin{align} 
(\mu_1^F-1) \alpha e^{-q^F} &= \int_{q_S}^{\infty} (\alpha e^{-q} - \mu_2^S e^{-2q} ) dq = \alpha e^{-q^S} - \mu_2^S e^{-2q^S}/2. \label{eq:exptarget1}\\ 
(\mu_2^F-1) (1- e^{-2q^F})/2 &= \int_{0}^{q^S} (e^{-2q} - \alpha \mu_1^S e^{-q}) dq = (1-e^{-2q^S})/2 - \mu_1^S \alpha(1- e^{-q^S}). \label{eq:exptarget2}
\end{align}

The equilibrium condition necessitates that 

\begin{align}
    \mu_1^F \alpha e^{-q^F} &= \mu_2^F e^{-2q^F}, \label{eq:exfp} \\
    \mu_1^S \alpha e^{-q^S} &= \mu_2^S e^{-2q^S}. \label{eq:expsp}
\end{align}

Therefore, we get equations relating $e^{-q^F}, e^{-q^S}$ to the multipliers. We plug these relations on the set of equations above, and also make the following substitution for legibility $\mu_1^S = x, \mu_1^F = y, \mu_2^S = z, \mu_2^F = w$.

\begin{align}
\frac{x}{y} &= 4 \label{eq:fpsp1}\\
\frac{z}{w} &= 1 + \frac{3}{2} \frac{(1-\frac{y^2 \alpha^2}{w^2})}{\frac{y^2 \alpha^2}{w^2}} \label{eq:fpsp2}\\
(y-1) \frac{\alpha^2 y}{w} &= \alpha^2 \frac{x}{z} - \frac{z \frac{x^2 \alpha^2}{z^2} }{2} \label{eq:fpsp3}\\ 
(w-1)(1-\frac{y^2 \alpha^2}{w^2})/2 &= (1-\frac{x^2\alpha^2}{z^2})/2 - x \alpha (1- \frac{x \alpha}{z}) \label{eq:fpsp4}
\end{align}

We additionally are only interested in equilibria where $x, z \geq 1$ since lower mulitpliers on the second price platform are weakly dominated.  

The revenue for the FP platform is 

$$ \int_{0}^{q^F} \mu_2^F e^{-2q} dq + \int_{q^F}^{\infty} \mu_1^F \alpha e^{-q} dq = \mu_2^F (1-e^{-2q^F})/2 + \mu_1^F \alpha e^{-q^F}.$$ 
    
In terms of the variables established above, this is 

$$ w (1-(\frac{y \alpha}{w})^2)/2 + \frac{y^2 \alpha^2}{w}.$$

The revenue of the SP platform is 

$$ \int_{0}^{q^S} \mu_1^S \alpha e^{-q} dq + \int_{q^S}^{\infty} \mu_2^S e^{-2q} dq = \mu_1^S \alpha (1-e^{-q^S}) + \mu_2^S e^{-2q^S}/2.$$ 

In terms of the variables established above, this is 

$$ x \alpha (1-\frac{x \alpha}{z}) + z \frac{x^2 \alpha^2}{2z^2}. $$
\end{proof}

We can now compute the payoff matrix of the platforms' game induced by the advertisers' subgame equilibrium multipliers. Using Claims~\ref{cl:expfpfp},~\ref{cl:expspsp} and~\ref{cl:expfpsp}:

  \begin{table}[h!]
  \centering
    \setlength{\extrarowheight}{8pt}
    \begin{tabular}{cc|c|c|}
      & \multicolumn{1}{c}{} & \multicolumn{2}{c}{Platform 2}\\
      & \multicolumn{1}{c}{} & \multicolumn{1}{c}{SPA}  & \multicolumn{1}{c}{FPA} \\\cline{3-4}
      \multirow{2}*{Platform 1}  & SPA & $\left((3-4\alpha)\alpha, (3-4\alpha)\alpha \right)$ & $( x \alpha -\frac{x^2 \alpha^2}{2z}),  w/2 (1+ \frac{y^2 \alpha^2}{2w})$ \\\cline{3-4}
      & FPA & $( x \alpha -\frac{x^2 \alpha^2}{2z}),  w/2 (1+ \frac{y^2 \alpha^2}{2w})$
      & $(\frac{\alpha^2+1}{2},\frac{\alpha^2+1}{2})$ \\\cline{3-4}
    \end{tabular}
    \caption{Payoff matrix for the platforms' game after solving for the advertisers' subgame equlibrium, for the valuatiosn described in Section~\ref{sec:expvsexp}.}
    \label{table:expgame}
  \end{table}

We now prove that for $\alpha \in (1/4, 1/2)    $, SP is a dominant strategy of the game defined by Table~\ref{table:expgame}.  

\begin{proof}[Proof of Theorem~\ref{thm:expequilibria}]
We change the variables and let $t=\frac{y\alpha}{w}$ and $u=\frac{x\alpha}{z}$. Then we have
\begin{align}
\text{\Cref{eq:fpsp1}} &\iff z=\frac{4wt}{u}    
\end{align}
We plug it into Equations~\ref{eq:fpsp2}-\ref{eq:fpsp4}:
\begin{align}
\text{\Cref{eq:fpsp2}} &\iff \frac{z}{w}=1+\frac{3}{2}\cdot \frac{1-t^2}{t^2} \iff u=\frac{8t^3}{3-t^2} \label{eq:proof_exponential_1}\\
\text{\Cref{eq:fpsp3}} &\iff (\frac{wt}{\alpha}-1)\alpha t=\alpha u -\frac{1}{2}zu^2 \iff wt^2-\alpha t=\alpha u -2uwt \iff w=\frac{\alpha(u+t)}{t^2+2ut} \label{eq:proof_exponential_2}\\
\text{\Cref{eq:fpsp4}} &\iff \frac{1}{2}(w-1)(1-t^2) = \frac{1}{2}(1-u^2) - uz(1-u) \\
&\iff (w-1)(1-t^2) = 1-u^2-8wt(1-u)\iff w=\frac{2-u^2-t^2}{1-t^2+8t-8tu} \label{eq:proof_exponential_3}
\end{align}

By Equations \ref{eq:proof_exponential_1}, \ref{eq:proof_exponential_2} and \ref{eq:proof_exponential_3}, we have
\begin{align}
    \alpha=\alpha(t)=\frac{(2-u^2-t^2)(t^2+2ut)}{(1-t^2+8t-8tu)(u+t)}=\frac{(2-(\frac{8t^3}{3-t^2})^2-t^2)(t^2+2t\cdot\frac{8t^3}{3-t^2})}{(1-t^2+8t-8t\cdot\frac{8t^3}{3-t^2})(\frac{8t^3}{3-t^2}+t)}.
\end{align}
Here we write $\alpha(t)$ as a function of $t$. By \Cref{eq:exfp}, $t=\frac{y\alpha}{w}=e^{-q^F}\in (0,1)$. Similarly, by \Cref{eq:expsp}, $u=\frac{x\alpha}{z}=e^{-q_S}\in (0,1)$. By \Cref{eq:proof_exponential_1}, we further have $\frac{8t^3}{3-t^2}<1$, which implies $t<0.7$. One can verify that when $t\in (0, 0.7)$, $\alpha(t)$ is an increasing function on $t$ (see \Cref{fig:proof_exp1}. As $\alpha(t)\in [1/4, 1/2]$, we have that $0.36<t<0.609$.

\begin{figure}[h!]
    \centering
    \includegraphics[width=0.35\textwidth]{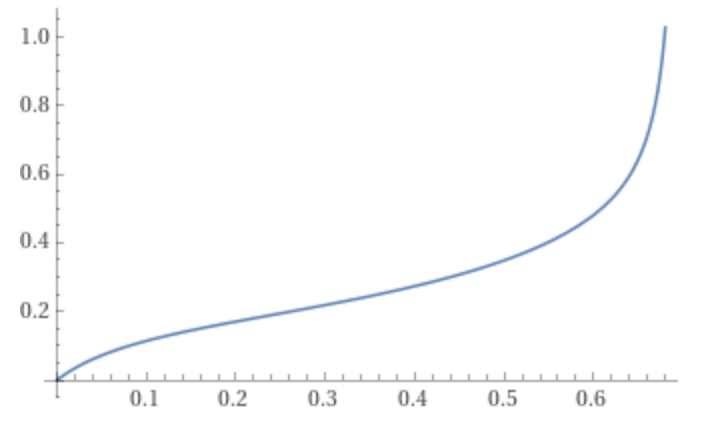}
    \caption{The function $\alpha(t)$ when $t\in [0, 0.7]$}
    \label{fig:proof_exp1}
\end{figure}


To show that SPA is a dominant strategy of the game, it suffices to show that
\begin{enumerate}
    \item $(3-4\alpha)\alpha >w/2 (1-(\frac{y \alpha}{w})^2) + \frac{y^2 \alpha^2}{w}$.
    \item $x \alpha (1-\frac{x \alpha}{z}) + \frac{x^2 \alpha^2}{2z} > \frac{\alpha^2+1}{2}$.
\end{enumerate}

We notice that when $\alpha\in [1/4, 1/2]$, $(3-4\alpha)\alpha \geq \frac{1}{2}$ and $\frac{\alpha^2+1}{2}\leq 5/8$. It's sufficient to show that (after rewriting the inequalities using $t, u$):
\begin{align}
    \frac{1}{2}>\frac{1}{2}w(1+t^2)\iff F_1(t)=\frac{2-(\frac{8t^3}{3-t^2})^2-t^2}{1-t^2+8t-8t\cdot \frac{8t^3}{3-t^2}}-\frac{1}{1+t^2}<0.\\
    4wt(1-\frac{u}{2})>\frac{5}{8}\iff F_2(t)=\frac{4(2-(\frac{8t^3}{3-t^2})^2-t^2)\cdot t\cdot (1-\frac{8t^3}{2(3-t^2)})}{1-t^2+8t-8t\cdot \frac{8t^3}{3-t^2}}-\frac{5}{8}>0.
\end{align}

One can easily verify that $F_2(t)>0>F_1(t)$ when $t\in (0.36, 0.609)$ (see \Cref{fig:proof_exp3} and \Cref{fig:proof_exp2}). Thus SPA is dominant strategy of the game.

\begin{figure}[h!]
    \centering
    \includegraphics[width=0.35\textwidth]{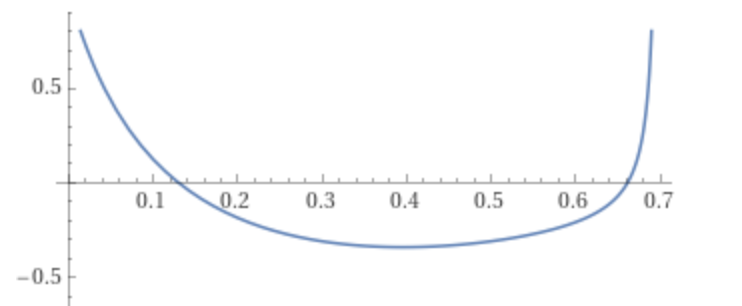}
    \caption{$F_1(t)$ when $t\in [0, 0.7]$}
    \label{fig:proof_exp3}
\end{figure}


\begin{figure}[h!]
    \centering
    \includegraphics[width=0.35\textwidth]{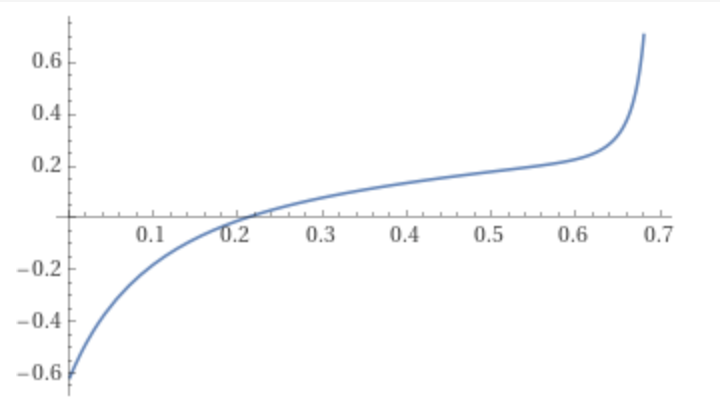}
    \caption{$F_2(t)$ when $t\in [0, 0.7]$}
    \label{fig:proof_exp2}
\end{figure}


\end{proof}

\end{document}